%% file: vldb.tex
\documentclass[sigconf, nonacm]{acmart}
\input{vldb_style}
\usepackage{units}
\usepackage{enumitem}
\usepackage{multirow}
\usepackage{subfigure}
\usepackage{graphicx}
\def \me{\texttt{MPE}}

\begin{document}
\title{Mixed-Precision Embeddings for Large-Scale Recommendation Models}
\input{sections/author}
\input{sections/abstract}
\maketitle
\input{vldb_block}

\input{sections/introduction}
\input{sections/preliminaries}
\input{sections/methodology}
\input{sections/experiments}
\input{sections/related_work}

\input{sections/conclusion}

\clearpage
\bibliographystyle{ACM-Reference-Format}
\bibliography{vldb}
\end{document}

%% file: vldb_style.tex
\newcommand\vldbdoi{XX.XX/XXX.XX}
\newcommand\vldbpages{XXX-XXX}
\newcommand\vldbvolume{14}
\newcommand\vldbissue{1}
\newcommand\vldbyear{2020}
\newcommand\vldbauthors{\authors}
\newcommand\vldbtitle{\shorttitle} 
\newcommand\vldbavailabilityurl{https://github.com/Leopold1423/mpe}
\newcommand\vldbpagestyle{plain} 

%% file: sections/author.tex
\author{Shiwei Li}
\authornote{This work was done when Shiwei Li worked as an intern at FiT, Tencent.}
\orcid{0000-0002-7067-0275}
\affiliation{
  \institution{Huazhong University of Science and Technology}
  \city{Wuhan}
  \country{China}
  \postcode{430074}
}
\email{lishiwei@hust.edu.cn}

\author{Zhuoqi Hu}
\orcid{0009-0006-9446-8465}
\affiliation{
  \institution{Huazhong University of Science and Technology}
  \city{Wuhan}
  \country{China}
  \postcode{430074}
}
\email{qzhycloud@outlook.com}


\author{Xing Tang}
\orcid{0000-0003-4360-0754}
\affiliation{
  \institution{FiT, Tencent}
  \city{Shenzhen}
  \country{China}
  \postcode{518054}
}
\email{xing.tang@hotmail.com}

\author{Haozhao Wang}
\orcid{0000-0002-7591-5315}
\affiliation{
  \institution{Huazhong University of Science and Technology}
  \city{Wuhan}
  \country{China}
  \postcode{430074}
}
\email{hz_wang@hust.edu.cn}

\author{Shijie Xu}
\orcid{0009-0000-6279-190X}
\affiliation{
  \institution{FiT, Tencent}
  \city{Shenzhen}
  \country{China}
  \postcode{518054}
}
\email{shijiexu@tencent.com}

\author{Weihong Luo}
\orcid{0009-0004-4329-4923}
\affiliation{
  \institution{FiT, Tencent}
  \city{Shenzhen}
  \country{China}
  \postcode{518054}
}
\email{lobbyluo@tencent.com}

\author{Yuhua Li}
\orcid{0000-0002-1846-4941}
\affiliation{
  \institution{Huazhong University of Science and Technology}
  \city{Wuhan}
  \country{China}
  \postcode{430074}
}
\email{idcliyuhua@hust.edu.cn}


\author{Xiuqiang He}
\orcid{0000-0002-4115-8205}
\affiliation{
  \institution{FiT, Tencent}
  \city{Shenzhen}
  \country{China}
  \postcode{518054}
}
\email{xiuqianghe@tencent.com}

\author{Ruixuan Li}
\authornote{Ruixuan Li is the corresponding author.}
\orcid{0000-0002-7791-5511}
\affiliation{
  \institution{Huazhong University of Science and Technology}
  \city{Wuhan}
  \country{China}
  \postcode{430074}
}
\email{rxli@hust.edu.cn}

%% file: sections/abstract.tex
\begin{abstract}
Embedding techniques have become essential components of large databases in the deep learning era. By encoding discrete entities, such as words, items, or graph nodes, into continuous vector spaces, embeddings facilitate more efficient storage, retrieval, and processing in large databases. Especially in the domain of recommender systems, millions of categorical features are encoded as unique embedding vectors, which facilitates the modeling of similarities and interactions among features. However, numerous embedding vectors can result in significant storage overhead. In this paper, we aim to compress the embedding table through quantization techniques. Given that features vary in importance levels, we seek to identify an appropriate precision for each feature to balance model accuracy and memory usage. To this end, we propose a novel embedding compression method, termed Mixed-Precision Embeddings (\me). Specifically, to reduce the size of the search space, we first group features by frequency and then search precision for each feature group. \me\ further learns the probability distribution over precision levels for each feature group, which can be used to identify the most suitable precision with a specially designed sampling strategy. Extensive experiments on three public datasets demonstrate that \me\ significantly outperforms existing embedding compression methods. Remarkably, \me\ achieves about 200x compression on the Criteo dataset without comprising the prediction accuracy.
\end{abstract}

%% file: vldb_block.tex
\pagestyle{\vldbpagestyle}
\begingroup\small\noindent\raggedright\textbf{PVLDB Reference Format:}\\
\vldbauthors. \vldbtitle. PVLDB, \vldbvolume(\vldbissue): \vldbpages, \vldbyear.\\
\href{https://doi.org/\vldbdoi}{doi:\vldbdoi}
\endgroup
\begingroup
\renewcommand\thefootnote{}\footnote{\noindent
This work is licensed under the Creative Commons BY-NC-ND 4.0 International License. Visit \url{https://creativecommons.org/licenses/by-nc-nd/4.0/} to view a copy of this license. For any use beyond those covered by this license, obtain permission by emailing \href{mailto:info@vldb.org}{info@vldb.org}. Copyright is held by the owner/author(s). Publication rights licensed to the VLDB Endowment. \\
\raggedright Proceedings of the VLDB Endowment, Vol. \vldbvolume, No. \vldbissue\ %
ISSN 2150-8097. \\
\href{https://doi.org/\vldbdoi}{doi:\vldbdoi} \\
}\addtocounter{footnote}{-1}\endgroup

\ifdefempty{\vldbavailabilityurl}{}{
\vspace{.3cm}
\begingroup\small\noindent\raggedright\textbf{PVLDB Artifact Availability:}\\
The source code, data, and/or other artifacts have been made available at \url{\vldbavailabilityurl}.
\endgroup
}

%% file: sections/introduction.tex
\section{Introduction}\label{sec:introduction}
\begin{figure*}[htbp]
    \centering
    \subfigure[The typical architecture of DLRMs.]{\label{fig:dlrms}
    \begin{minipage}[h]{0.685\textwidth}
    \centering
    \includegraphics[scale=0.96]{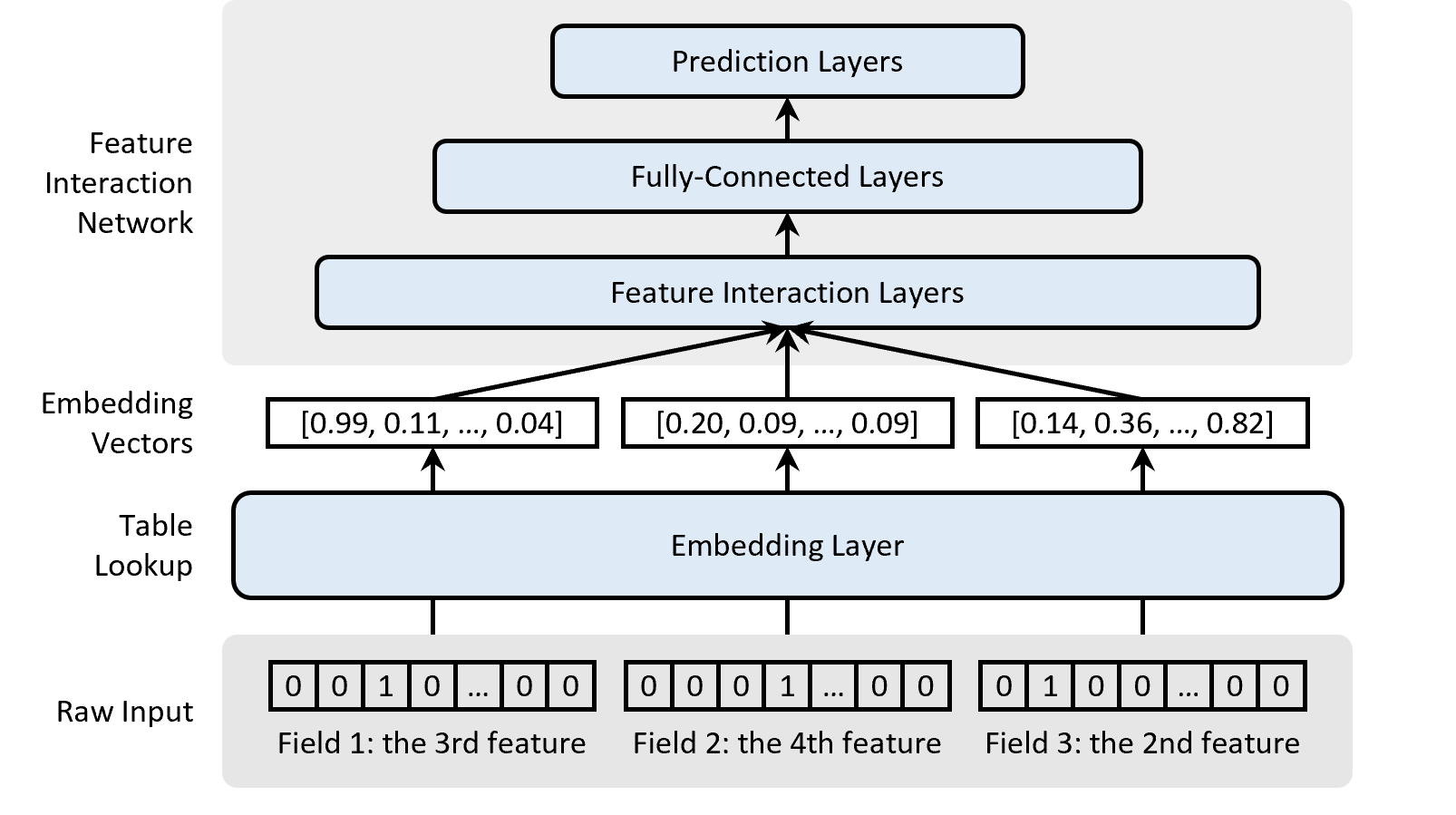}
    \end{minipage}
    }
    \subfigure[The schematic diagram of QAT.]{\label{fig:qat}
    \begin{minipage}[h]{0.285\textwidth}
    \centering
    \includegraphics[scale=0.96]{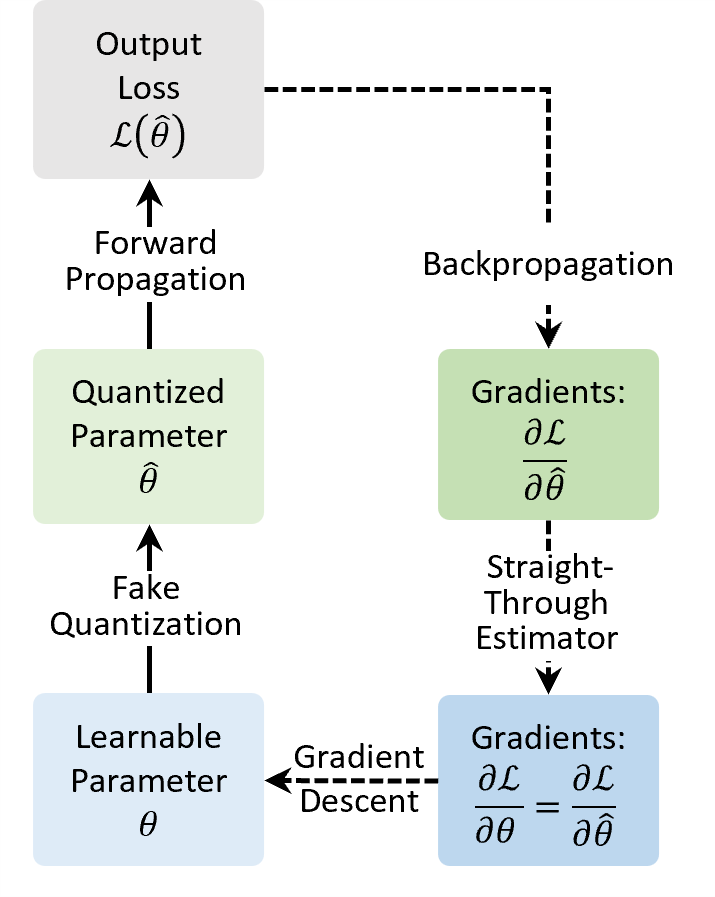}
    \end{minipage}
    }    
    \caption{The typical architecture of deep learning recommendation models (DLRMs) and the schematic diagram of quantization-aware training (QAT). (a) DLRMs are usually composed of an embedding layer and a feature interaction network. (b) A fake quantizer will be inserted into the forward propagation of QAT, and end-to-end optimization is then achieved through Straight-Through Estimator (STE), which treats the quantizer as an identity map during backpropagation.}
    \Description{The typical architecture of deep learning recommendation models (DLRMs) and the schematic diagram of quantization-aware training (QAT).}
    \label{fig:dlrms_and_qat}
\end{figure*}

\subsection{Background}
In the deep learning era, embedding techniques have become essential components of large databases due to their effectiveness and efficiency in representing and retrieving information \cite{database-sigmoid,database-cikm}. By encoding discrete entities, such as words, items, or graph nodes, into continuous vector spaces, embeddings facilitate more efficient storage, retrieval, and processing within large databases. These techniques are now widely applied across various domains, including natural language processing \cite{word2vec,embedding-nlp-survey}, graph learning \cite{node2vec,embedding-graph-icml}, and recommender systems \cite{embed_survey,ctr-survey,generalized-embedding}. 
In particular, embedding techniques have a great impact on recommender systems, where the number of required embeddings far exceeds that of other domains. 
Specifically, the input data in recommender systems typically consists of various attributes, such as user\_id, age, and occupation, each with multiple possible values. These attributes are often referred to as feature fields, while their specific values are known as categorical features or simply features. Each categorical feature will be represented by a unique embedding vector, which enhances the modeling of similarities and interactions among features, thereby improving the accuracy of recommendations. 
However, the amount of categorical features in recommender systems is significantly larger than that of discrete entities in other domains. For example, an ordinary advertising system in Baidu contains billions of features~\cite{baidu_lpt}. In contrast, the vocabulary size of the large language model LLaMA 3 \cite{llama3} is only 128,000, with the former being several orders of magnitude larger. Therefore, embedding techniques are particularly important in recommender systems.

The embedding techniques are commonly adopted as part of the deep learning recommendation models (DLRMs).
The typical architecture of DLRMs includes an embedding table that converts categorical features into dense vector representations, followed by a feature interaction network that processes these embeddings to generate predictions, as illustrated in Figure \ref{fig:dlrms}. Numerous studies have refined the structure of the feature interaction network to improve prediction accuracy \cite{deepfm, dcn, ipnn, dcnv2, optinter, afm}. Despite these improvements, the interaction networks often have relatively shallow layers and a limited number of parameters, while the majority of model parameters are concentrated in the embedding table. Especially when handling billions of categorical features, the storage of the embedding table can even reach 10 TB \cite{baidu_lpt}. Therefore, developing techniques to compress embeddings while maintaining model accuracy has become a crucial issue in deploying DLRMs.


Regarding model compression, quantization is undoubtedly one of the most effective techniques \cite{white-paper-quantization}, which reduces memory usage by representing parameters with fewer bits. There are two primary methods of quantization: post-training quantization (PTQ) \cite{ptq_embedding,adaround} and quantization-aware training (QAT) \cite{lsq,lsq_plus,qat-oscillations}. PTQ applies quantization after the training process, while QAT incorporates a quantization operator during training. As shown in Figure \ref{fig:qat}, QAT inserts a fake quantization operator into the forward propagation, allowing the model to better adapt to quantization and thus achieve higher accuracy than PTQ. During backpropagation, the learnable parameters are updated using the Straight-Through Estimator (STE) \cite{ste}, approximating the gradient of the non-differentiable quantization operator as 1, thereby enabling end-to-end optimization. After training, the parameters will be quantized and then stored in a low-precision format for deployment.

\subsection{Motivation and Contributions}
An embedding table can be represented by a matrix $\boldsymbol{E} \in \mathbb{R}^{n\times d}$, where $n$ is the number of categorical features, and $d$ is the dimension of each embedding vector. As discussed by \citet{cafe}, existing methods generally compress either the rows or columns of the embedding table. Row compression methods involve techniques like feature selection \cite{optfs,adaembed,lpfs,erase} and hashing \cite{qr-trick,double-hash,binary-hash,cafe}. For example, OptFS \cite{optfs} filters out unimportant features to reduce the number of rows in the embedding table, while \citet{double-hash} propose sharing embeddings among infrequent features through hash functions, thereby reusing rows. On the other hand, dimension compression methods involve techniques like mixed-dimension embeddings \cite{cprec,autoemb,autodim,optembed} and pruning \cite{pep,deeplight,ham}. For example, CpRec \cite{cprec} assigns varying dimensions to embeddings of different features, while PEP \cite{pep} prunes redundant parameters within each feature embedding by learning a threshold. A key characteristic of these methods is their ability to explicitly or implicitly assess feature importance and allocate storage space accordingly.

In addition to row and column perspectives, parameter precision is another critical perspective of embedding compression, closely related to quantization techniques. \citet{ptq_embedding} were the first to apply PTQ to embedding tables. Later studies \cite{baidu_lpt,facebook_lpt,alpt} have focused on the low-precision training (LPT) paradigm. Unlike QAT, which retains full-precision parameters during training, LPT keeps embeddings in a low-precision format throughout training to reduce training memory usage. However, at lower precision levels, LPT becomes impractical due to significant accuracy loss. For example, ALPT \cite{alpt} achieves lossless compression only at 8-bit precision. Note that recommender systems typically cannot tolerate any degradation in accuracy. Beyond these efforts, there has been little in-depth research on quantization, particularly QAT, within recommender systems. This scarcity is largely attributable to the unique characteristic of embedding tables in recommender systems, which render quantization techniques less prevalent compared to domains such as computer vision \cite{lsq} and natural language processing \cite{quantized-lstm}. 
Specifically, embedding tables often contain many features with varying levels of importance. Yet, quantization typically applies uniform precision across all embeddings, failing to distinguish feature importance. As a result, numerous less important features are assigned the same precision as more significant ones, even though they could be represented with lower precision, leading to inefficient compression. 

In this paper, we aim to improve the quantization of embedding tables by incorporating the ability to distinguish feature importance. Previous studies have linked feature importance to the embedding dimension \cite{cprec}, the utilization of the feature \cite{optfs} and so on. Similarly, we propose to associate feature importance with embedding precision. More precisely, our objective is to find the most appropriate precision for each feature embedding to balance model accuracy with memory efficiency. It is worth noting that feature selection can be viewed as a binary case of mixed-precision embeddings, where the search space includes only zero-precision and full-precision options. Therefore, mixed-precision embeddings can achieve finer-grained compression than feature selection methods.

To solve the problem of searching embedding precision, we propose a novel algorithm named Mixed-Precision Embeddings (\me). There are two main challenges in building \me. 
The first challenge is the vast number of categorical features. DLRMs typically involve millions of categorical features, resulting in an excessively large search space when searching precision for each feature individually. To mitigate this issue, we propose grouping features and optimizing precision for each group, substantially reducing search space complexity. Previous research has identified feature frequency as a reliable measure of feature importance \cite{double-hash,adaembed}. Therefore, we suggest grouping features by frequency to ensure that features within the same group have similar levels of importance, thereby minimizing conflicts from shared precision. 
The second challenge is determining each feature group's precision (i.e., quantization bit-width). However, precision search is inherently a discrete problem that traditional gradient descent algorithms cannot solve effectively. To achieve end-to-end optimization, we propose learning a probability distribution over candidate bit-widths for each group, where output embeddings will be represented as the expected values of quantized embeddings across all candidate bit-widths. Additionally, we apply regularization to the expected bit-width of groups to minimize memory usage. By balancing the objective function with this regularization, we can effectively learn the optimal probability distribution over candidate bit-widths. Finally, we sample the precision based on the optimized probability distribution and then retrain the model to enhance accuracy.

Our main contributions can be summarized as follows:
\begin{itemize}[leftmargin=*]
\item We identify a critical limitation of quantization-aware training in embedding compression for recommendation models, namely, its inability to distinguish feature importance.
\item We propose a novel embedding compression method, \me, that can implicitly distinguish feature importance. Specifically, \me\ identifies an appropriate precision for each feature embedding.
\item We evaluate \me\ on three large-scale public datasets across various models. The experimental results demonstrate \me\ significantly outperforms existing embedding compression methods. 
\end{itemize}

%% file: sections/preliminaries.tex
\section{Preliminaries}
\def \cb{\mathbb{B}}
\def \ci{\mathcal{I}}
\def \cq{\mathcal{Q}}
\def \cd{\mathbb{D}}
\def \ce{\mathcal{E}}
\def \cf{\mathcal{F}}
\def \cl{\mathcal{L}}
\def \e{\boldsymbol{e}}
\def \p{\boldsymbol{p}}
\def \E{\boldsymbol{E}}
\def \B{\boldsymbol{B}}
\def \W{\boldsymbol{W}}

In this section, we provide a formal definition of deep learning recommendation models and quantization-aware training.

\subsection{Deep Learning Recommendation Models}\label{sec:dlrms}
Figure \ref{fig:dlrms} illustrates the general architecture of DLRMs, which includes an embedding layer $\ce$ and a feature interaction network $\cf$. For simplicity, we use $\E$ and $\W$ to denote the parameters within the embedding layer and the feature interaction network, respectively. The embedding layer maps categorical features $x$ into dense embeddings, represented as $\ce(x) = \ci(x)^T \E$, where $x$ is the input features and $\ci(x)$ is the corresponding one-hot encoded vectors as shown in Figure \ref{fig:dlrms}. These embeddings are then processed by the following network to generate predictions. Denoting the dataset by $\cd$ and the loss function by $\cl$, the optimization process of DLRMs can be formulated as follows:
\begin{equation}\label{eq:dlrms}
    \min_{\E, \W} \sum_{(x, y)\in \cd}\cl(y, \cf(\ce(x, \E), \W))
\end{equation}
The commonly used notations are summarized in Table \ref{tab:notation}. 

\input{tables/notation}

\subsection{Quantization-Aware Training}\label{sec:qat}
In the context of neural networks, quantization is to reduce the precision of parameters. By converting full-precision weights into lower-precision formats, such as 8-bit integers, quantization can significantly decrease the memory usage of neural networks. 
Consequently, we seek to reduce the storage overhead of embedding tables through quantization. It is worth noting that we do not quantize the feature interaction network, as doing so would severely hurt prediction accuracy while yielding little memory savings.

In this paper, we employ the uniform quantization \cite{white-paper-quantization} to compress embedding tables. Specifically, given a step size $\alpha$, an offset $\beta$, and the quantization bit-width $b$, a full-precision number $\theta$ will be quantized as follows:
\begin{equation}\label{eq:quantization}
\begin{aligned}
    \hat{\theta} = \cq(\theta,\alpha,\beta, b) = \alpha \bar{\theta} + \beta, \\
    \bar{\theta} = \text{clamp}(\lfloor \frac{\theta-\beta}{\alpha} \rceil, N_b, P_b), \\
    N_b = -2^{m-1}, \quad P_b = 2^{m-1}-1, \\
\end{aligned}
\end{equation}
where $\hat{\theta}$ and $\bar{\theta}$ are the quantized value and corresponding integer value, respectively. $N_b$ and $P_b$ are the negative and positive bounds of a $b$-bit signed integer, respectively. 
$\lfloor x \rceil$ rounds $x$ to its nearest integer and the function ${\text{clamp}}(\cdot)$ ensures that the returned integer stays within $[N_b, P_b]$. The step size $\alpha$ determines the granularity of the quantization, and the offset $\beta$ is the center of the quantized values. Note that $\alpha$ and $\beta$ are shared across lots of parameters, such as those within each layer, making their storage negligible. 
As illustrated in Figure \ref{fig:qat}, the above quantizer $\cq$ can be inserted into a network to facilitate QAT. When applying $b$-bit QAT to embedding tables, the training objective of DLRMs can be formulated as follows:
\begin{equation}\label{eq:qat-dlrms}
    \min_{\E, \alpha, \beta, \W} \sum_{(x, y)\in \cd}\cl(y, \cf(\ce(x, \cq(\E, \alpha, \beta, b)), \W))
\end{equation}

However, the non-differentiable nature of the quantizer $\cq$, particularly due to the rounding function, presents challenges for optimization with gradient descent algorithms. Previous work has utilized the Straight-through Estimator (STE) \cite{ste} to approximate the gradient of the quantizer as 1. In this paper, we employ an advanced QAT algorithm, LSQ+ \cite{lsq_plus}, as the base quantizer, which similarly approximates the gradient of the rounding function as 1. This allows the gradients for the input parameters $\theta$, $\alpha$, and $\beta$ to be calculated, as detailed below:

\begin{equation}\label{eq:grad_theta}
\frac{\partial \hat{\theta}}{\partial \theta} 
= \frac{\partial \bar{\theta}}{\partial \theta} \alpha
= \left\{
\begin{array}{ll}
    1   & \quad \text{if} \  N_b<\frac{\theta-\beta}{\alpha}<P_b, \\ 
    0   & \quad \text{otherwise}.
\end{array}
\right.
\end{equation}

\begin{equation}\label{eq:grad_alpha}
\frac{\partial \hat{\theta}}{\partial \alpha} 
= \frac{\partial \bar{\theta}}{\partial \alpha} \alpha + \bar{\theta} 
= \left\{
\begin{array}{ll}
    N_b     & \quad \text{if} \  \frac{\theta-\beta}{\alpha}\leq N_b, \\  
    -\frac{\theta-\beta}{\alpha} + \lfloor \frac{\theta-\beta}{\alpha} \rceil 
    & \quad \text{if} \  N_b<\frac{\theta-\beta}{\alpha}<P_b, \\ 
    P_b     & \quad \text{if} \  \frac{\theta-\beta}{\alpha}\geq P_b. \\
\end{array}
\right.
\end{equation}

\begin{equation}\label{eq:grad_beta}
\frac{\partial \hat{\theta}}{\partial \beta} 
= \frac{\partial \bar{\theta}}{\partial \beta} \alpha +1 
= \left\{
\begin{array}{ll}
    0   & \quad \text{if} \  N_b<\frac{\theta-\beta}{\alpha}<P_b \\ 
    1   & \quad \text{otherwise}.
\end{array}
\right.
\end{equation}

%% file: tables/notation.tex
\begin{table}[!htbp]
\centering
\caption{Commonly used notations and descriptions.}
\resizebox{0.48\textwidth}{!}{
\begin{tabular}{cl}
\toprule 
\textbf{Notation} & \textbf{Description} \\ \toprule
$\cd$   & Dataset. \\ 
$\cl$   & Loss function. \\ 
$\ce$   & Function of the embedding layer. \\ 
$\cf$   & Function of the feature interaction network. \\ 
$\E$    & Parameters within the embedding table. \\
$\W$    & Parameters within the feature interaction network. \\
$n, d$  & Number of features and embedding dimension. \\
$x, y$  & Input features $x$ and the label $y$ of a data sample. \\
$\cq$   & Quantizer. \\ 
$\cb$   & Set of candidate bit-widths. \\ 
$m$         & Number of candidate bit-widths. \\ 
$b, \alpha, \beta$  & Bit-width, step size, and offset of quantization. \\ 
$N_b, P_b$          & Negative and positive bounds of $b$-bit signed integers. \\ 
$\e, \hat{\e}$      & Full-precision and quantized embedding vector. \\ 
$g$         & Number of feature groups. \\ 
$s^k$       & Sum of the feature frequencies in the $k$-th group. \\ 
$\gamma$    & Learnable probability distribution parameters. \\ 
$\tau$      & Temperature coefficient in the softmax function. \\ 
$p$         & Probability distribution over precision levels. \\ 
$\lambda$   & Regularization coefficient of expected bit-width. \\ 
\bottomrule
\end{tabular}
}
\label{tab:notation}
\end{table}

%% file: sections/methodology.tex
\section{Mixed-Precision Embeddings}
In this section, we first provide a formal definition of embedding precision search problem in Section \ref{sec:problem}. Then, we introduce how to reduce the complexity of the search space in Section \ref{sec:grouping} and how to search precision for feature embedding in Section \ref{sec:probability}. Finally, we illustrate how to generate mixed-precision embedding with the searched precision in Section \ref{sec:retrain}.

\begin{figure*}
    \centering
    \includegraphics[width=0.87\linewidth]{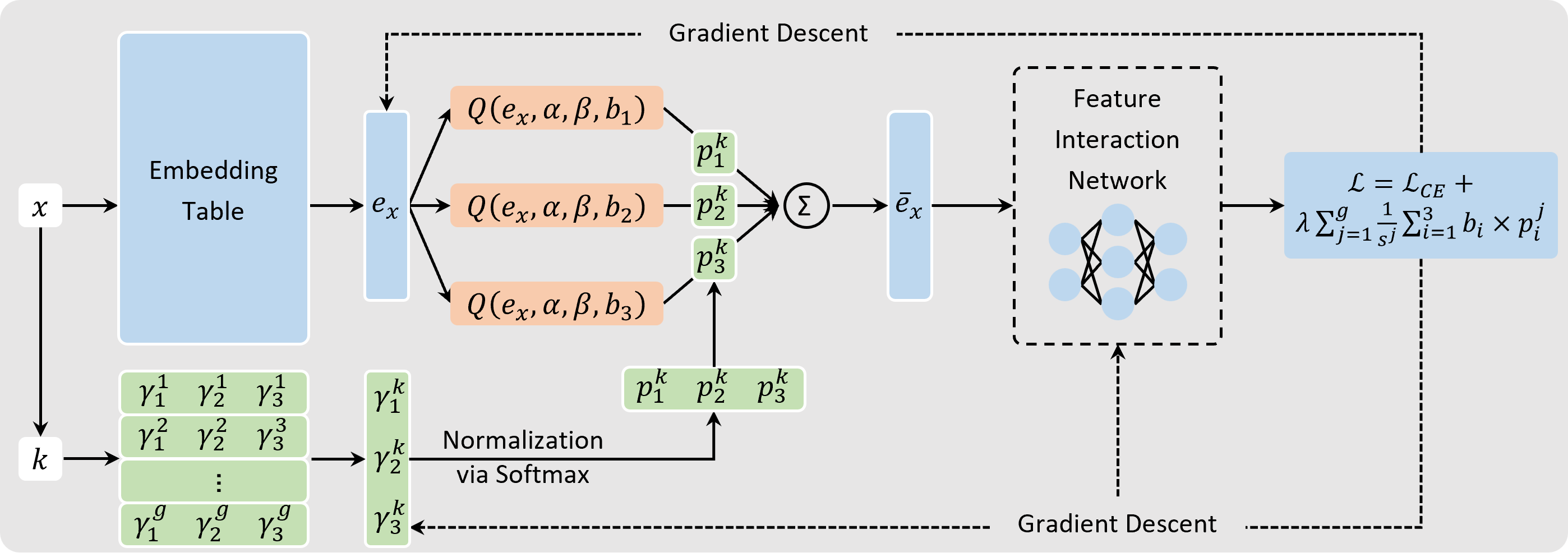}
    \caption{Learning process of the probability distribution over candidate bit-widths in \me. $x$ is a feature of the input data, and $k$ is the corresponding group index when sorted by feature frequency.}
    \Description{Learning process of the probability distribution over candidate bit-widths in \me.}
    \label{fig:mpe}
\end{figure*}

\subsection{Problem Formulation}\label{sec:problem}
It is a consensus that categorical features in recommender systems have varying levels of importance \cite{double-hash,optfs}. Previous studies leveraged this variation in feature importance to reduce the parameters associated with less important features \cite{double-hash,autoemb,optfs,adaembed}. For example, \citet{cprec} reduces the embedding dimension of less important features through heuristic rules, while AutoEmb~\cite{autoemb} automatically searches embedding dimensions with reinforcement learning algorithms. Additionally, AdaEmbed~\cite{adaembed} uses gradient information and feature frequencies to access feature importance and then prunes unimportant features. Despite these advancements, no prior work has systematically combined feature importance with corresponding embedding precision. In this paper, to achieve more efficient embedding compression, we propose searching for an optimal precision level for each feature, which refers to the smallest bit-width without losing prediction accuracy.

Given a set of candidate precision levels (i.e., bit-widths), $\cb = \{b_1, b_2, ..., b_m\}$, we aim to determine $\B \in \cb^{n \times 1}$ for the embedding table $\E \in \mathbb{R}^{n \times d}$, where each element of $\B$ represents the precision of the corresponding embedding in $\E$. Specifically, the problem of mixed-precision embedding can be formally defined as follows:
\begin{equation}\label{eq:mpe-dlrms}
\begin{aligned}
    \min_{\B, \E, \alpha, \beta, \W} \sum_{(x, y)\in \cd}\cl(y, \cf(\ce(x, \cq(\E, \alpha, \beta, \B)), \W)), \\ s.t. \quad \Vert\B\Vert_1 \leq n\times B^*,
\end{aligned}
\end{equation}
where $B^*$ is the average bit-width budget and $\B$ is included as a parameter of $\cq$ to signify mixed-precision quantization. Notably, the step size $\alpha$ and the offset $\beta$ are still shared across numerous parameters. Additionally, when $0$ is included in $\cb$ and assigned to a feature, the corresponding embedding is set to a zero vector. In this case, it also serves as a feature selection mechanism.


\subsection{Frequency-Aware Grouping}\label{sec:grouping}
Due to the large number of features, the search space for embedding precision becomes vast, complicating the optimization of Eq.(\ref{eq:mpe-dlrms}). However, since the number of candidate bit-widths is significantly smaller than the number of features, many features can share the same bit-width. Therefore, we can group features in advance and ensure that features in the same group can satisfy the same bit-width. Feature frequency, a common measure of feature importance used in numerous studies \cite{cprec, adaembed, double-hash, mgqe}, is employed here as prior information to guide the grouping process. Specifically, features are sorted by frequency and divided into $g$ groups, with a bit-width searched for each group individually. This simple strategy reduces the search space complexity by a factor of $\nicefrac{n}{g}$ while minimizing conflicts arising from the shared bit-width within a group.

\subsection{Learnable Probability Distribution}\label{sec:probability}
In the previous section, we reduced the search space by grouping features. The remaining challenge is determining the optimal precision for each feature group. 
However, optimizing the embedding precision is more complex than solving discrete problems like feature selection or embedding dimension search. Feature selection is a simple decision about whether to include a feature while embedding dimension search involves selecting from multiple candidate dimensions. In contrast, mixed-precision embedding presents two discrete optimization problems: selecting from multiple candidate bit-widths and managing non-differentiable quantization operators.

Inspired by traditional QAT algorithms, we employ the STE to handle the non-differentiability of quantization operators. Further, to optimize the precision of different feature groups in an end-to-end manner, we convert the one-hot selection of candidate bit-widths into the learning of a probability distribution. Specifically, for a given set of candidate bit-widths, $\cb=\{b_1, b_2, ..., b_m\}$, we aim to learn a probability distribution $\p=\{p_1, p_2, ..., p_m\}$ for each feature group, where $p_i$ denotes the likelihood of selecting $b_i$ as the bit-width. To ensure the probabilities sum to 1, we maintain a learnable vector $\gamma=\{\gamma_1, \gamma_2, ..., \gamma_m\}$ for each group and calculate the probabilities using the softmax function, as shown below:
\begin{equation}
p_i = \frac{e^{\nicefrac{\gamma_i}{\tau}}}{\sum_{j=1}^m e^{\nicefrac{\gamma_j}{\tau}}},
\end{equation}
where $\tau$ is the temperature used to control the optimization of $\gamma$, and a smaller $\tau$ brings $\p$ closer to a one-hot distribution. Initially, all $\gamma$ values are set to zero, ensuring that each candidate bit-width has an equal probability when training begins. When necessary, superscript $k$ will be used for $\p$ and $\gamma$ to indicate their association with the $k$-th group. 

Using the probability distribution above, each feature in a group has a specific probability of being quantized at various bit-widths. Therefore, we use the expected outcomes of quantizing the feature embedding at different bit-widths as the final quantized embedding, as shown in Figure \ref{fig:mpe}. Specifically, for an embedding vector $\e$, the final quantized embedding $\bar{\e}$ will be computed as follows:
\begin{equation}
    \bar{\e} = \sum_{i=1}^m p_i \times \cq(\e, \alpha, \beta, b_i),
\end{equation}
where $\cq$ is the base quantizer, LSQ+ \cite{lsq_plus}, as discussed in Section \ref{sec:qat}. 
The quantized embedding $\bar{\e}$ is then fed into the subsequent networks to compute the loss function. During backpropagation, the original embedding $\e$ and the probability distribution parameter $\gamma$ will be updated through gradient descent algorithms.

To balance the storage overhead of embedding tables with model prediction accuracy, we introduce regularization for the expected bit-width of each feature group. Given the correlation between feature importance and frequency, we utilize the sum of the feature frequencies within each group to adjust the regularization coefficients accordingly. Specifically, the objective function we optimize is defined as follows:
\begin{equation}
    \cl = \cl_{CE} + \lambda \sum_{j=1}^g \frac{1}{s^j} \sum_{i=1}^m b_i \times p_i^j
\end{equation}
where $\cl_{CE}$ represents the cross-entropy loss calculated between the model outputs and the ground-truth labels, $s^j$ is the sum of feature frequencies in the $j$-th group, and $\lambda$ is the regularization coefficient used to balance model accuracy and memory usage.

Additionally, we emphasize that maintaining separate step size and offset for each feature is both memory-intensive and unnecessary. Instead, quantization with different bit-widths necessitates distinct step sizes, while varying dimensions of the embedding table require different offsets. Therefore, we adopt a single step size for each bit-width and a single offset for each embedding dimension.

\subsection{Precision Sampling and Retraining}\label{sec:retrain}
Through the optimization process illustrated in Figure \ref{fig:mpe}, we can derive the probability distribution over candidate bit-widths, allowing us to determine each group's final bit-width. However, the bit-width with the highest probability may not necessarily be the most appropriate choice.
Specifically, while a higher bit-width may not correspond to the highest probability, it can still contribute a considerable portion of high-precision representation to the final embedding. In this context, this feature group should be recognized as one that necessitates a higher bit-width. To this end, we select the highest bit-width with a probability exceeding the threshold of $\frac{1}{2m}$ as the final precision for each group, as outlined below:
\begin{equation}
    b^*=\max\{b_i \ | \ p_i>\frac{1}{2m}, i\in[1, 2, ..., m]\},
\end{equation}
where $b^*$ is the sampled precision of a feature group. Bit-widths with a probability less than $\frac{1}{2m}$ are considered to have minimal contribution to the high-precision representation of embeddings.

On the other hand, directly quantizing the embedding parameters using the sampled bit-width may create discrepancies between the quantized embedding and that generated during training, thus affecting the model's accuracy. This discrepancy arises because the embedding used during training result from combining quantization outcomes across all candidate bit-widths. To mitigate accuracy loss, we implement a retraining process using the sampled bit-widths to train mixed-precision embedding from scratch. Following the Lottery Ticket Hypothesis (LTH) \cite{lth}, previous studies \cite{pep,autodim} will reset the model parameters to the initialization values for retraining. 
However, to leverage the information from the search phase, we employ the optimized step size, offset, and feature interaction network to initialize the model in the retraining phase, aiming to enhance training outcomes. Note that only the embeddings are set back to the initialized values of the search phase during retraining.

\section{Implementation}
\me\ is implemented as a plug-in embedding layer module based on PyTorch, which can be seamlessly integrated into existing recommendation models with minimal modifications. There are two stages in the implementation of \me, training and inference. During training, embeddings are maintained as full-precision parameters, while during inference, they are supposed to be stored in low-precision formats. To enhance usability, the training and inference stages are encapsulated in separate modules. 

During inference, \me\ only requires storing embedding in mixed-precision formats and does not involve mixed-precision computations. Once the low-precision embeddings are retrieved, they must be dequantized into full-precision parameters before being fed into subsequent networks. Note that the input to the embedding table is highly sparse, and the embedding retrieved during forward propagation constitutes only a small proportion of the total. 
As a result, the storage overhead for dequantized embedding parameters is negligible. However, since frameworks such as PyTorch do not natively support data formats such as Int-2 and Int-3, we concatenate each embedding vector at the bit level and then divide them into the Int-16 format for storage. Upon feature retrieval, the corresponding Int-16 data will be converted into full-precision embedding. 

%% file: sections/experiments.tex
\section{Experiments}
In this section, we aim to demonstrate the superiority of \me\ in embedding compression for DLRMs through extensive experiments, addressing the following research questions:
\begin{itemize}
    \item \textbf{RQ1:} How does \me\ compare to state-of-the-art methods in terms of compression efficiency for embedding tables?
    \item \textbf{RQ2:} How does retraining affect the performance of \me?
    \item \textbf{RQ3:} Are the bit-widths optimized for a specific model applicable to other models?
    \item \textbf{RQ4:} How does \me\ perform in terms of inference latency?
    \item \textbf{RQ5:} Is \me\ capable of generating mixed-precision embeddings?
\end{itemize}

\subsection{Experimental Settings}
\subsubsection{Datasets.}
The experiments are conducted on three real-world datasets: Avazu \cite{avazu}, Criteo \cite{criteo}, and KDD12 \cite{kdd12}. The Avazu dataset contains 10 days of click logs with 22 categorical feature fields. The Criteo dataset contains 7 days of ad click data with 26 categorical and 13 numerical feature fields. The KDD12 dataset has no temporal information and consists of 11 categorical feature fields. The statistics of the datasets are shown in Table \ref{tab:datasets}.

Following previous work~\cite{alpt,optfs}, we preprocess the datasets as outlined below. For the Avazu dataset, we remove the instance\_id field and transform the timestamp field into three new fields: hour, weekday, and is\_weekend. For the Criteo dataset, each numeric value $x$ is discretized into $\lfloor \log^2(x)\rfloor$ when $x > 2$; otherwise, $x$ is set to 1. For all datasets, categorical features that appear only once are replaced with an "OOV" (out-of-vocabulary) token. Each dataset is then randomly split in the ratio of 8:1:1 to obtain the corresponding training, validation, and test sets.

\begin{table}[!htbp]
\renewcommand{\arraystretch}{1}
\centering
\caption{Statistics of the datasets.}\label{tab:datasets}
\resizebox{0.47\textwidth}{!}{
\begin{tabular}{ccccc}
\toprule[1pt]
\textbf{Dataset} & \textbf{\#Fields} & \textbf{\#Features} & \textbf{\#Samples} & \textbf{Positive Ratio} \\ 
\toprule[1pt]
Avazu            & 22 & 9,449,445  & 40,428,967  & 16.98\% \\ 
Criteo           & 39 & 33,762,577 & 45,840,617  & 25.62\% \\ 
KDD12            & 11 & 54,689,798 & 149,639,105 & 4.45\% \\ 
\bottomrule[1pt]
\end{tabular}}
\end{table}

\subsubsection{Models.} 
In the experiments, we evaluated four widely-used models: DNN, DCN \cite{dcn}, DeepFM \cite{deepfm}, and IPNN \cite{ipnn}. The DNN model employs a multi-layer perception (MLP) as the feature interaction network to process embeddings and generate predictions. In contrast, the other three models integrate additional modules for feature interaction modeling. 
In addition to the MLP, the DCN, DeepFM, and IPNN models integrate a cross network, a factorization machine (FM) \cite{fm}, and an inner product layer, respectively. 
These four classic models represent different patterns of feature interaction and are chosen to demonstrate the versatility of \me.

\input{tables/overall_table}

\subsubsection{Baselines.}
We compare the proposed \me\ with various embedding compression methods, including QR-Trick \cite{qr-trick}, ALPT \cite{alpt}, LSQ+ \cite{lsq_plus}, PEP \cite{pep}, OptFS \cite{optfs}. QR-Trick is a hashing method, using multiple hash functions and complementary embedding tables to reduce the collisions among features. ALPT is an embedding quantization method, specifically a low-precision training method, which compresses both training and inference memory usage. LSQ+ is an advanced QAT method designed to quantize the weights and activations of convolutional neural networks with a specific bit-width. In this paper, we employ LSQ+ to compress embedding tables of DLRMs. PEP is to prune the feature embeddings with a learnable threshold. OptFS is a feature selection method that learns a gate scalar to decide whether to keep a feature. The implementations of all baselines strictly refer to the original papers or codes, which are available in the open source code repository.

\subsubsection{Metrics.} 
We assess the prediction accuracy of various methods on the test set using AUC (Area Under the ROC Curve) and Logloss (cross-entropy loss), where a higher AUC or lower Logloss indicates better recommendation performance. Note that a 0.001 increase in AUC is generally considered a significant improvement. To assess compression efficiency, we present the embedding storage ratios relative to uncompressed embedding tables across different methods. Additionally, we discuss inference latency in Section \ref{sec:inference}.

\subsubsection{Hyperparameters.}
We set various hyperparameters following previous work to ensure that our backbone has sufficiently excellent performance. Specifically, we set the embedding dimension to 16 and initialize the embedding parameters by a normal distribution with a standard deviation of 3e-3. For all four models, we use a three-layer MLP where the size of the fully-connect layers are 1024, 512, 256, respectively. Adam \cite{adam} is employed as the optimizer. The batch size is set to 10000 and Batch Normalization \cite{bn} is used to ensure stable training. The learning rate is carefully tuned among $\{$1e-4, 3e-4, 1e-3, 3e-3, 1e-2$\}$ and the weight decay is tuned among $\{${0.0, 1e-7, 3e-7, 1e-6, 3e-6, 1e-5}$\}$. It turns out that the optimal learning rate is 1e-3, the optimal weight decay coefficient of the Avazu and KDD12 datasets is 0.0, and that of the Criteo dataset is 3e-6. Additionally, both the Avazu and KDD12 datasets reached optimal accuracy at the end of the first epoch, and overfitting occurred immediately after entering the second epoch. Therefore, we train one epoch on the Avazu and KDD12 datasets while training 8 epochs on the Criteo dataset. 
For \me, we set the temperature coefficient $\tau$ to 3e-3 and the group size to 128. These two parameters require little tuning. The set of candidate bit-widths is set to $\{0,1,2,3,4,5,6\}$, where the maximum bit width is determined by the performance of LSQ+. Specifically, LSQ+ achieves a minimum bit-width of 6 without sacrificing prediction accuracy on all three datasets. We mainly adjust the regularization coefficient $\lambda$ among $\{$1e-6, 3e-6, 1e-5, 3e-5, 1e-4, 3e-4$\}$ for \me\ to achieve a trade-off between model accuracy and compression ratio. For more detailed implementation, please refer to the open-source code. 
Each experiment is run at least five times on Nvidia 3090 GPUs with Intel Xeon E5-2673 CPUs. Average results and the standard deviation are reported.

\subsection{Overall Performance (RQ1)}\label{sec:overall}
In this section, we primarily address two questions: (1) the maximum compression strength of various methods without losing accuracy, and (2) how the accuracy of these methods varies with different compression ratios. Specifically, we regulate the compression ratio for each method by adjusting relevant hyper-parameters, such as the bit-width for LSQ+ and ALPT. However, the QR-trick achieves a minimum compression ratio of 2x and cannot provide lossless compression, making it an exception.

\begin{figure}[t]
\centering
\subfigure[Avazu]{
\begin{minipage}[h]{0.49\textwidth}
\centering
\includegraphics[width=0.9\linewidth]{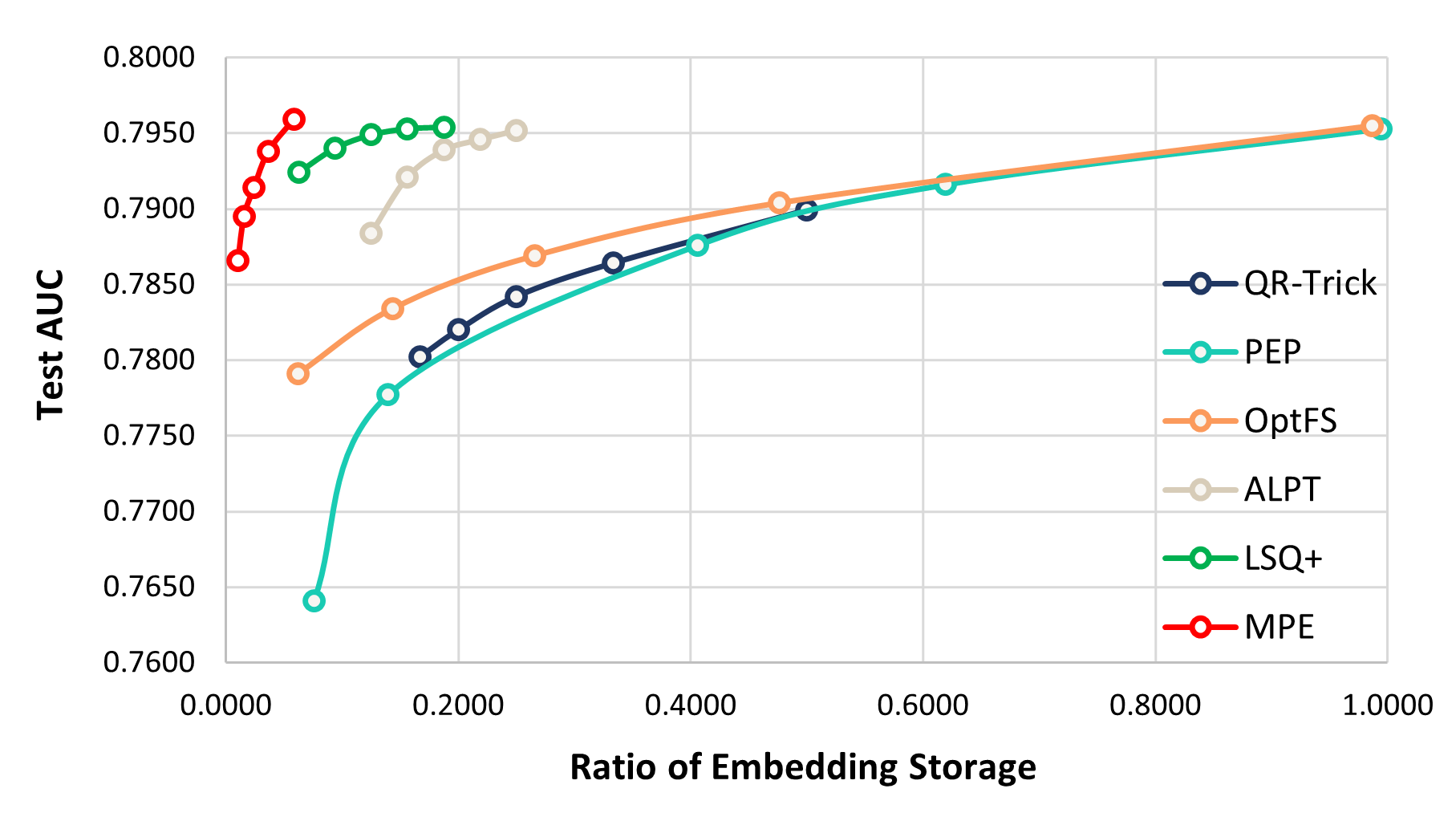}
\end{minipage}
}
\subfigure[Criteo]{
\begin{minipage}[h]{0.49\textwidth}
\centering
\includegraphics[width=0.9\linewidth]{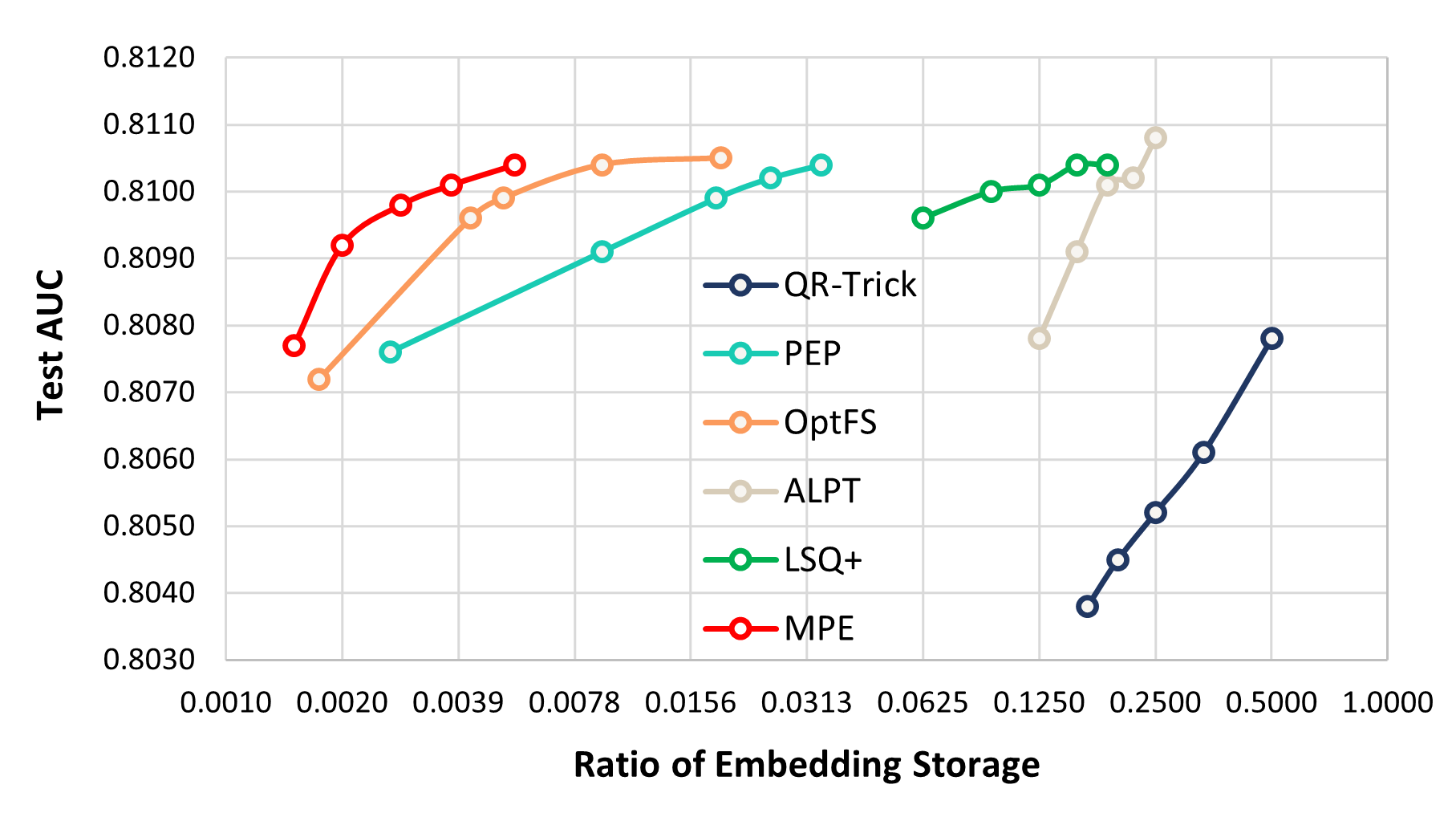}
\end{minipage}
}
\subfigure[KDD12]{
\begin{minipage}[h]{0.49\textwidth}
\centering
\includegraphics[width=0.9\linewidth]{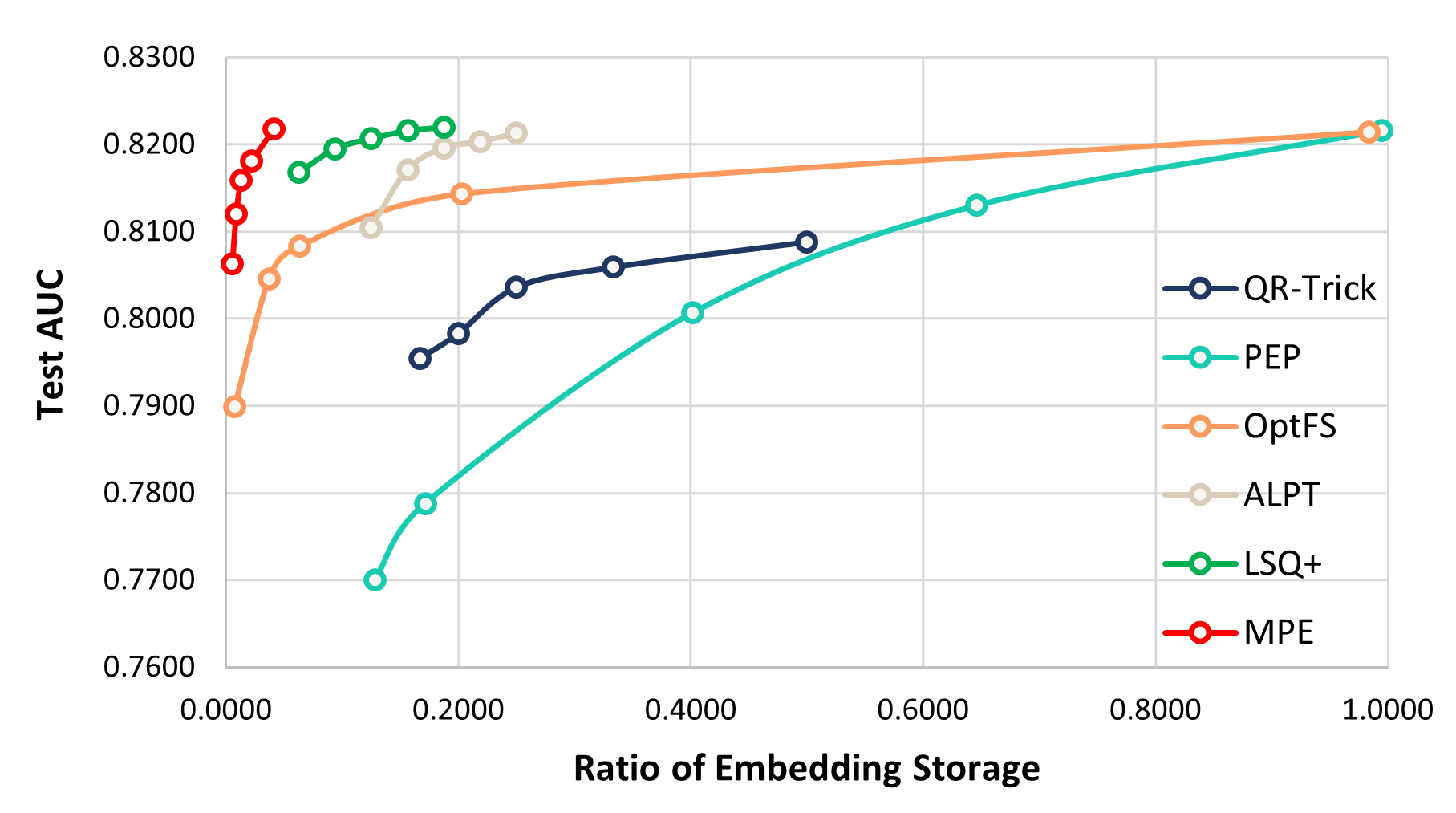}
\end{minipage}
}
\Description{AUC under varying compression ratios.}
\caption{AUC under varying compression ratios.}\label{fig:ratio}
\end{figure}

Table \ref{tab:overall} presents the accuracy and compression ratios of various methods at lossless compression, which demonstrates that the compression efficiency of \me\ significantly surpasses that of other baselines. Specifically, \me\ achieves approximately 18, 200, and 20 times compression on embedding tables without comprising prediction accuracy on the three datasets, where the average bit widths are 1.8, 0.16, and 1.6, respectively. 
The QR-trick shares embeddings across features, which diminishes feature discrimination when multiple features use the same representation. Even with just a 2x compression, the QR-trick leads to significant accuracy loss. 
LSQ+ achieves a minimum bit-width of 6 without compromising accuracy on all three datasets. Due to the lack of full-precision parameter support during training, ALPT requires a minimum bit-width of 8, which is slightly higher than that of LSQ+. 
From the pruning perspective, PEP and OptFS prune the embedding table at the element-level and row-level, respectively. They have some compression under the Criteo dataset, while the compression under the Avazu and KDD12 datasets is almost negligible, suggesting that there are few useless features or redundant parameters under the latter two datasets. 
However, LSQ+ can still compress embeddings into 6-bit integer parameters under the Avazu and KDD12 datasets, which indicates that the redundancy in the bit-width of parameters is significantly higher than that in the number of parameters. In contrast, on the Criteo dataset, the compression performance of LSQ+ is inferior to OptFS and PEP, primarily due to the uniform bit-width applied by LSQ+. Even when some features could be represented with fewer bits, they must use the same bit-width as more important features. 
Compared to LSQ+, \me\ achieves finer-grained compression through the proposed mixed-precision mechanism, assigning different bit-widths to different features. Compared to OptFS and PEP, \me\ can also distinguish the importance of different features and achieves more efficient compression by reducing the precision of embedding parameters. Figure \ref{fig:ratio} illustrates the accuracy of various methods across different compression ratios, employing the DNN model. Obviously, at the same compression ratio, \me\ consistently outperforms all baseline methods in terms of the AUC.

\subsection{Ablation on Retraining (RQ2)}\label{sec:ablation}
In this section, we examine the impact of retraining on \me. We evaluated two variants: one without retraining and another with retraining using the LTH, where all parameters are reset to their initial values from the search phase. As shown in Table \ref{tab:retraining}, quantizing embeddings with the sampled bit-widths without retraining leads to a significant loss of accuracy. However, retraining with LTH can partially mitigate this loss, although some degradation in accuracy remains. In \me, we use the best-performing parameters, except for embeddings, to improve the initialization of the retrained model, resulting in substantial gains in accuracy. Moreover, the impact of retraining varies across datasets, with minimal effects on the Criteo dataset and more significant effects on the other two datasets. 

\begin{table}[!ht]
\renewcommand\arraystretch{1.1}
\centering
\caption{Ablation on retraining with the DNN model.}\label{tab:retraining}
\resizebox{0.47\textwidth}{!}{
\begin{tabular}{ccccccc}
\toprule
\multirow{2}{*}{\textbf{retraining}} & \multicolumn{2}{c}{\textbf{Avazu}}  & \multicolumn{2}{c}{\textbf{Criteo}} & \multicolumn{2}{c}{\textbf{KDD12}} \\
\cmidrule{2-7}
& AUC & LogLoss & AUC & LogLoss & AUC & LogLoss \\
\midrule
w.o. & 0.7919 & 0.3728  & 0.8102 & 0.4415  & 0.8190 & 0.1483 \\
LTH  & 0.7930 & 0.3722  & 0.8105 & 0.4413  & 0.8204 & 0.1477 \\
\me & \textbf{0.7959} & \textbf{0.3702}  & \textbf{0.8104} & \textbf{0.4415}  & \textbf{0.8218} & \textbf{0.1475} \\
\bottomrule
\end{tabular}
}
\end{table}

\begin{figure*}[!ht]
\centering
\subfigure[Avazu]{
\begin{minipage}[h]{0.3\textwidth}
\centering
\includegraphics[width=0.8\linewidth]{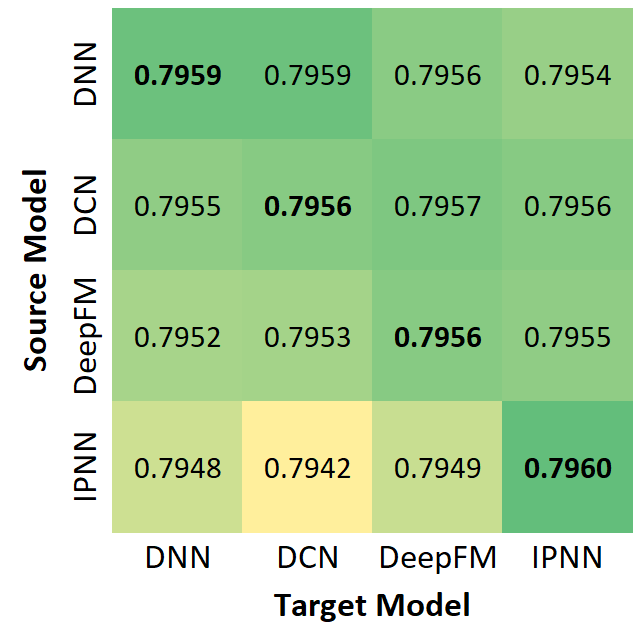}
\end{minipage}
}
\subfigure[Criteo]{
\begin{minipage}[h]{0.3\textwidth}
\centering
\includegraphics[width=0.8\linewidth]{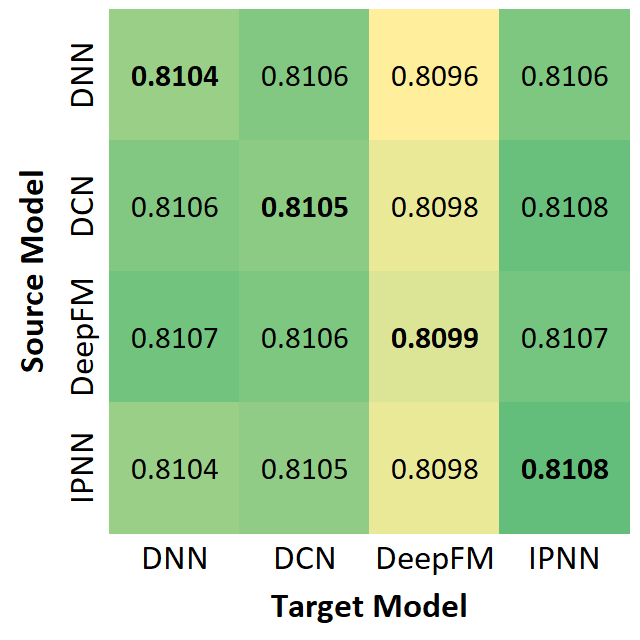}
\end{minipage}
}
\subfigure[KDD12]{
\begin{minipage}[h]{0.3\textwidth}
\centering
\includegraphics[width=0.8\linewidth]{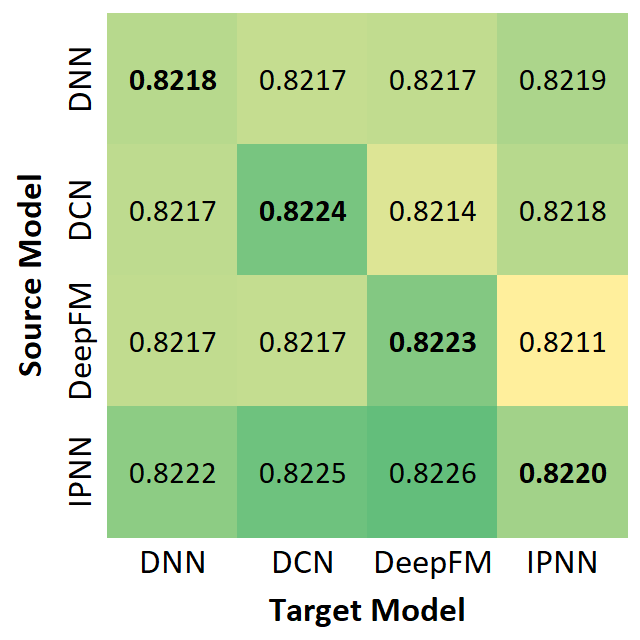}
\end{minipage}
}
\Description{Transferability analysis.}
\caption{Transferability analysis. Each column contains the test AUC of a specific target model with different source models. }\label{fig:transfer}
\end{figure*}

\begin{figure*}[!ht]
\centering
\subfigure[Avazu]{
\begin{minipage}[h]{0.3\textwidth}
\centering
\includegraphics[width=1.05\linewidth]{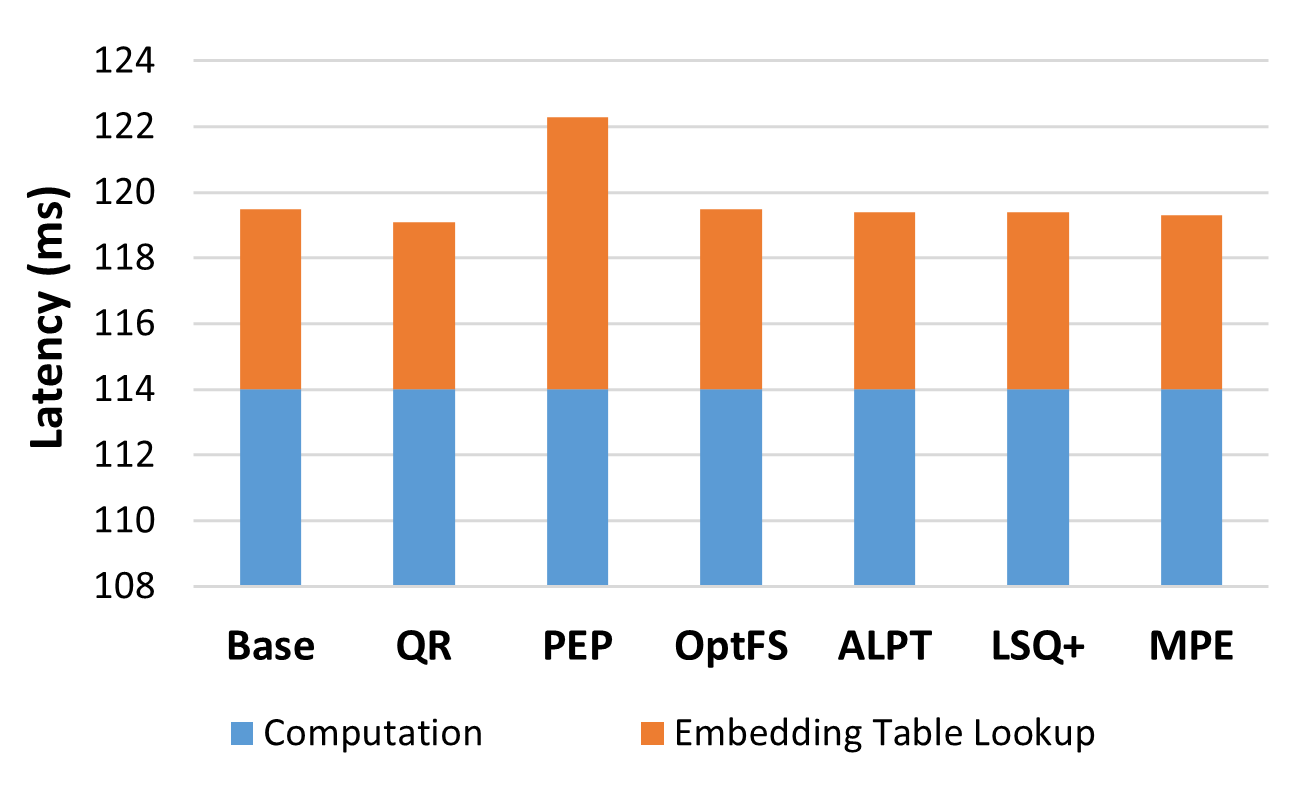}
\end{minipage}
}
\subfigure[Criteo]{
\begin{minipage}[h]{0.3\textwidth}
\centering
\includegraphics[width=1.05\linewidth]{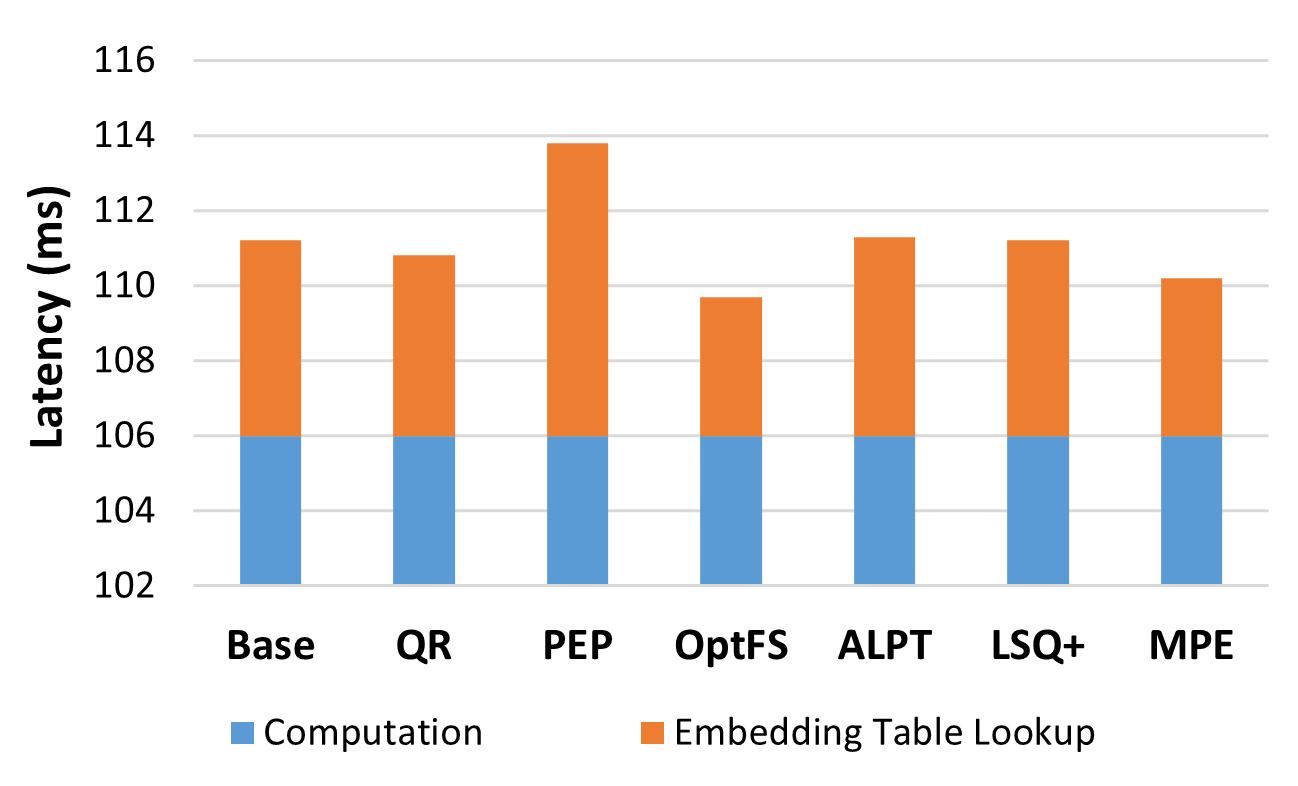}
\end{minipage}
}
\subfigure[KDD12]{
\begin{minipage}[h]{0.3\textwidth}
\centering
\includegraphics[width=1.05\linewidth]{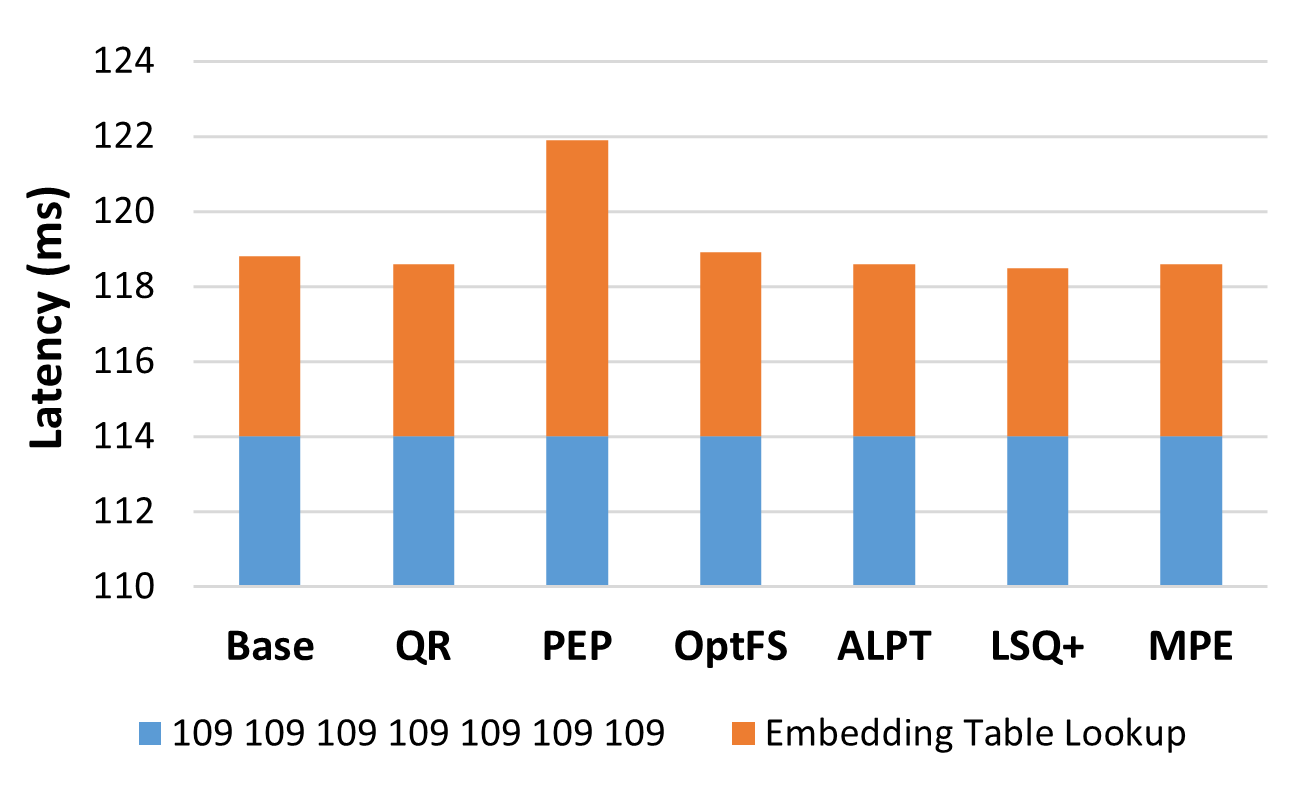}
\end{minipage}
}
\Description{Inference latency of the DNN model using different compression methods.}
\caption{Inference latency of the DNN model using different compression methods.}\label{fig:latency}
\end{figure*}

\subsection{Transferability (RQ3)}\label{sec:transfer}
In this section, we analyze the transferability of \me, which specifically refers to the performance of applying the sampled bit-widths from a specific model to other models for retraining. It is important to note that \me\ utilizes the step size, offset, and feature interaction network from the search phase to initialize the model during retraining. However, the structure of feature interaction networks may differ across models. Thus, we only reuse overlapping parameters. The models involved in the search and retraining processes are referred to as the source and target models, respectively.

Figure \ref{fig:transfer} presents the test AUC of the target model using various source models. To evaluate transferability, comparisons should be made within each column, ensuring that the target model remains consistent. On the Criteo dataset, sampled bit-widths can be transferred between different model architectures without impacting retraining accuracy. This aligns with the results in Table \ref{tab:retraining}, which indicate that retraining has little impact on the Criteo dataset. In contrast, for the Avazu and KDD12 datasets, transferring search results across architectures occasionally reduces model accuracy. This implies that feature precision under these datasets is more sensitive to the structure of the feature interaction network. Moreover, the reduction in accuracy can be partially attributed to the inability to fully reuse the parameters of the feature interaction network generated during the search phase. 
However, compared to the accuracy loss observed without retraining, the loss incurred from transferring search results between different models is negligible.

\subsection{Inference Latency (RQ4)}\label{sec:inference}
In this section, we evaluate the inference latency of different methods using the DNN model. The batch size used for inference is 10000. The primary difference among the methods lies in the embedding layer, and the time consumed by subsequent computations remain consistent. Therefore, we represent the inference latency as two parts: embedding table lookup and computation. As illustrated in Figure \ref{fig:latency}, the time spent on table lookup is significantly shorter than that on data loading and computation, resulting in no substantial difference in overall inference latency across methods. Notably, PEP requires sparse matrix storage for embedding compression, which adds latency to the table lookup process. OptFS reduces the number of features, especially for the Criteo dataset, thereby accelerating table lookup. Methods such as ALPT, LSQ+, and \me\ reduce the storage overhead of embedding vectors, speeding up table lookup. However, the dequantization process of low-precision data slightly offsets the acceleration effect.

\subsection{Bit-width Adjustment Capability (RQ5)}\label{sec:selection}
\begin{figure*}[t]
\centering
\subfigure[Avazu]{
\begin{minipage}[h]{0.3\textwidth}
\centering
\includegraphics[width=1.05\linewidth]{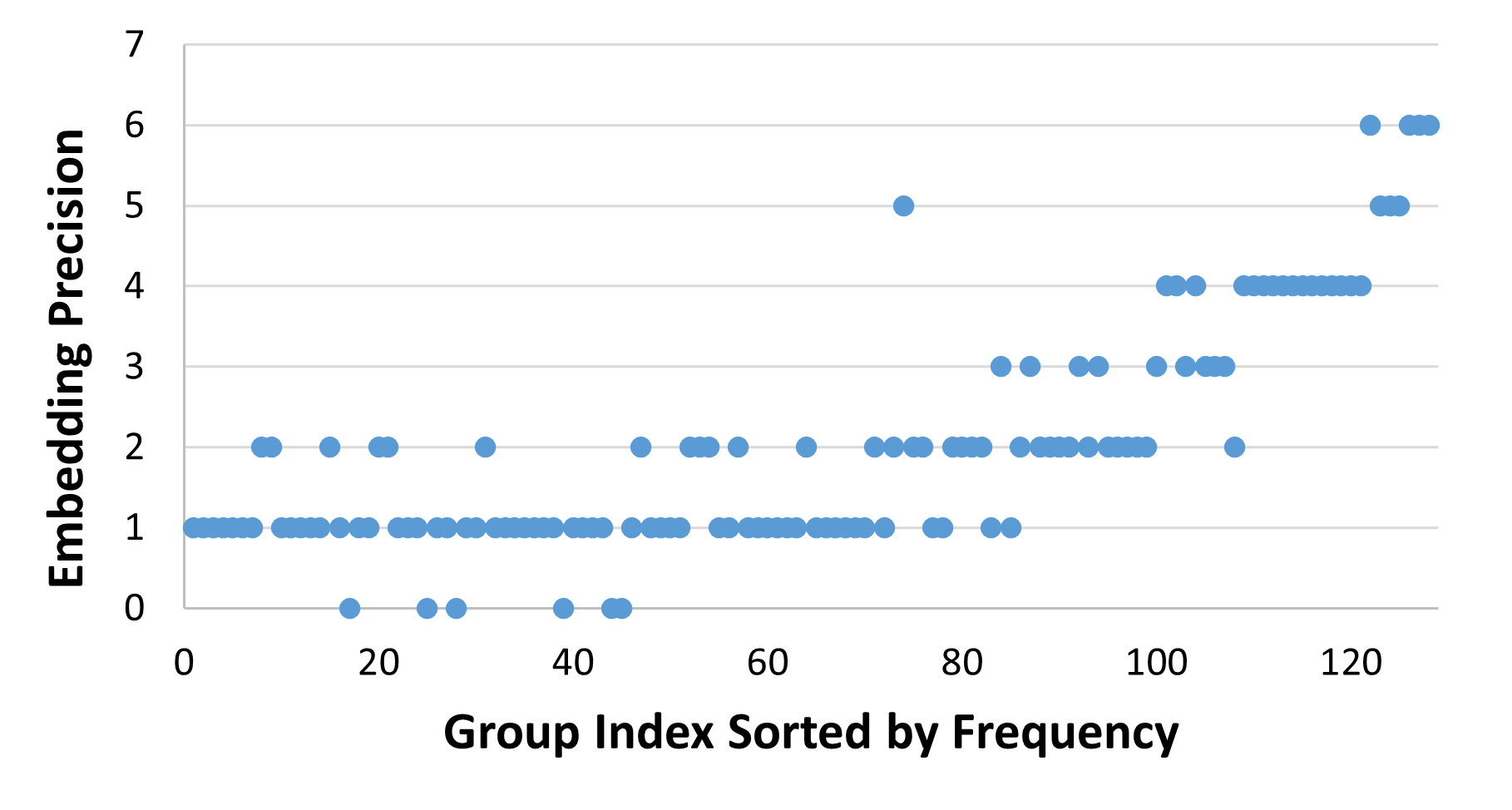}
\end{minipage}
}
\subfigure[Criteo]{
\begin{minipage}[h]{0.3\textwidth}
\centering
\includegraphics[width=1.05\linewidth]{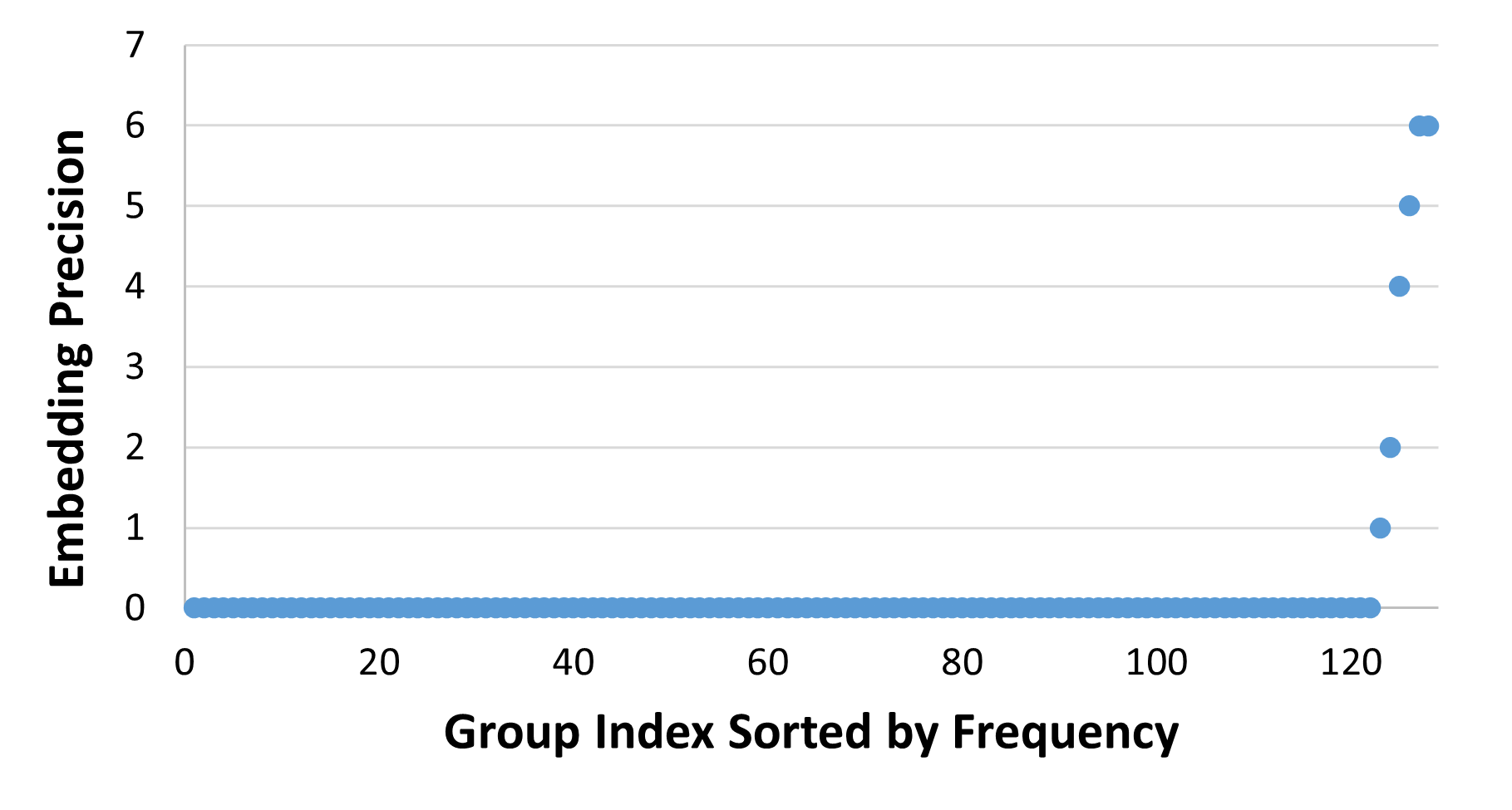}
\end{minipage}
}
\subfigure[KDD12]{
\begin{minipage}[h]{0.3\textwidth}
\centering
\includegraphics[width=1.05\linewidth]{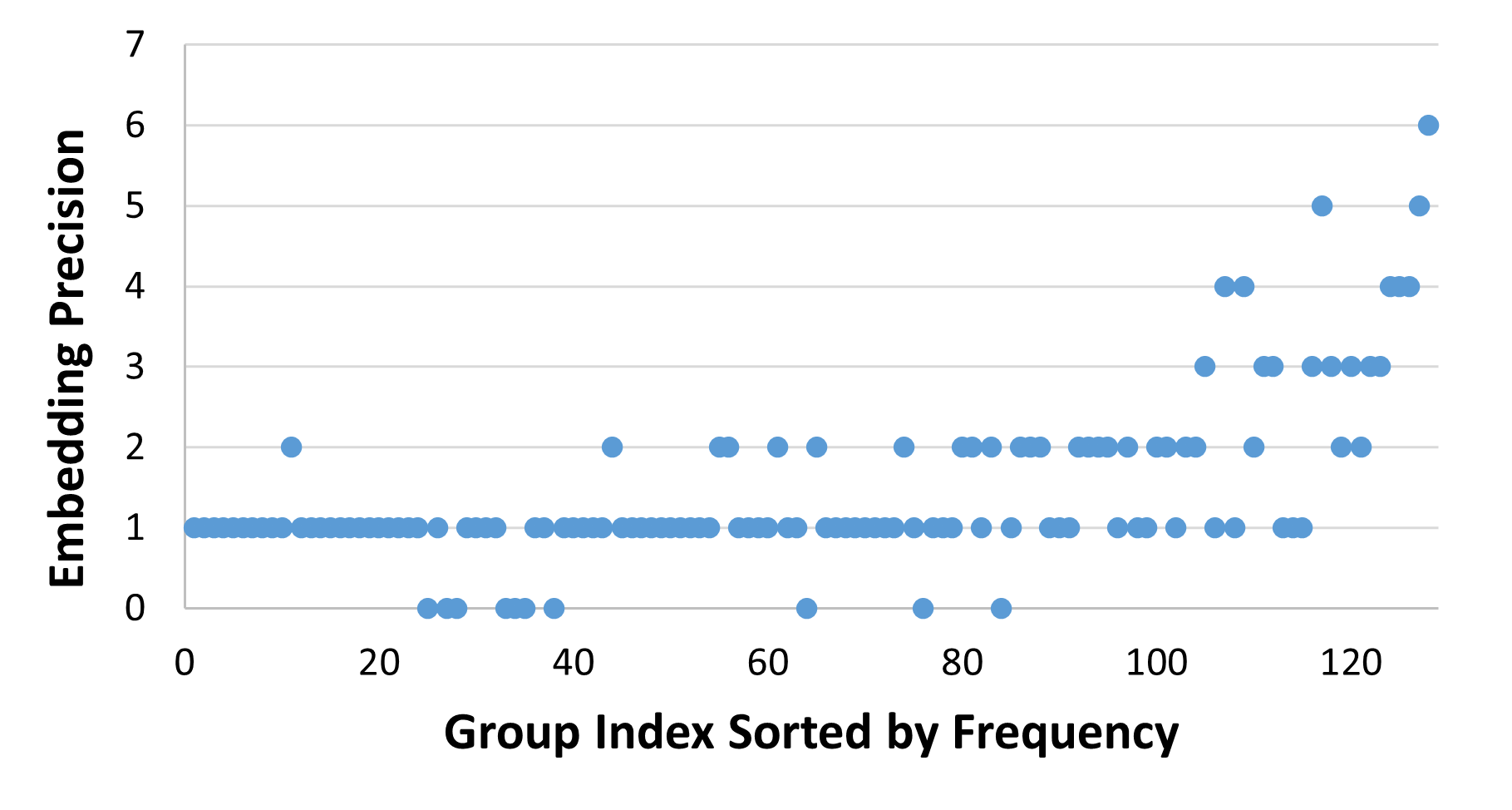}
\end{minipage}
}
\Description{Embedding precision of different feature groups.}
\caption{Embedding precision of different feature groups.}\label{fig:precision}
\end{figure*}

In this section, we provide a detailed examination of the sampled bit-widths. 
The candidate bit-widths of \me\ range from $\{$0, 1, 2, 3, 4, 5, 6$\}$. Assigning a bit-width of 0 to a feature produces a zero vector for the corresponding embedding, which is equivalent to performing feature selection.  As shown in Figure \ref{fig:precision}, the Avazu and KDD12 datasets exhibit minimal redundancy in features, whereas the Criteo dataset displays a greater degree of redundancy, consistent with the experimental results of OptFS. 
Additionally, Figure \ref{fig:precision} demonstrates that \me\ can dynamically adjust bit-widths across different feature groups. The sampled bit-widths further indicate a positive correlation between feature precision and frequency.

%% file: tables/overall_table.tex
\begin{table*}[!htbp]
\renewcommand\arraystretch{1.08}
\centering
\caption{Performance comparison between \me\ and baseline methods.}	\label{tab:overall}
\vspace{-5pt}
\resizebox{1.0\textwidth}{!}{
\begin{tabular}{c|c|ccc|ccc|ccc}
\hline
& \multirow{2}{*}{\textbf{Method}} & \multicolumn{3}{c}{\textbf{Avazu}} & \multicolumn{3}{|c}{\textbf{Criteo}} & \multicolumn{3}{|c}{\textbf{KDD12}}  \\ 
\cline{3-11}
& & AUC$\uparrow$ & Logloss$\downarrow$ & Ratio$\downarrow$ & AUC$\uparrow$ & Logloss$\downarrow$ & Ratio$\downarrow$ & AUC$\uparrow$ & Logloss$\downarrow$ & Ratio$\downarrow$  \\
\hline
\multirow{7}{*}{\rotatebox{90}{\textbf{DNN}}}
& Backbone & 0.7955 ($\pm$ 1e-4) & 0.3704 ($\pm$ 1e-4) & 1.0000 ($\pm$ 0e+0) & \textbf{0.8106} ($\pm$ 2e-4) & \textbf{0.4413} ($\pm$ 2e-4) & 1.0000 ($\pm$ 0e+0) & 0.8219 ($\pm$ 2e-4) & \textbf{0.1473} ($\pm$ 1e-4) & 1.0000 ($\pm$ 0e+0) \\
& QR-Trick & 0.7899 ($\pm$ 2e-4) & 0.3741 ($\pm$ 1e-4) & 0.5000 ($\pm$ 0e+0) & 0.8078 ($\pm$ 1e-4) & 0.4437 ($\pm$ 1e-4) & 0.5000 ($\pm$ 0e+0) & 0.8088 ($\pm$ 2e-4) & 0.1502 ($\pm$ 1e-4) & 0.5000 ($\pm$ 0e+0) \\
& PEP & 0.7953 ($\pm$ 1e-4) & 0.3705 ($\pm$ 1e-4) & 0.9945 ($\pm$ 1e-5) & 0.8104 ($\pm$ 1e-4) & 0.4415 ($\pm$ 1e-4) & 0.0340 ($\pm$ 1e-3) & 0.8216 ($\pm$ 1e-4) & 0.1475 ($\pm$ 1e-4) & 0.9947 ($\pm$ 2e-4) \\
& OptFS & 0.7955 ($\pm$ 1e-4) & 0.3705 ($\pm$ 1e-4) & 0.9863 ($\pm$ 2e-4) & 0.8105 ($\pm$ 1e-4) & 0.4414 ($\pm$ 1e-4) & 0.0187 ($\pm$ 2e-4) & 0.8214 ($\pm$ 1e-4) & 0.1475 ($\pm$ 1e-4) & 0.9839 ($\pm$ 1e-4) \\
& ALPT & 0.7952 ($\pm$ 1e-4) & 0.3707 ($\pm$ 1e-4) & 0.2500 ($\pm$ 0e+0) & 0.8104 ($\pm$ 1e-4) & \textbf{0.4413} ($\pm$ 1e-4) & 0.2500 ($\pm$ 0e+0) & 0.8214 ($\pm$ 1e-4) & 0.1474 ($\pm$ 1e-4) & 0.2500 ($\pm$ 0e+0) \\
& LSQ+ & 0.7954 ($\pm$ 1e-4) & 0.3705 ($\pm$ 1e-4) & 0.1875 ($\pm$ 0e+0) & 0.8104 ($\pm$ 1e-4) & 0.4414 ($\pm$ 1e-4) & 0.1875 ($\pm$ 0e+0) & \textbf{0.8220} ($\pm$ 1e-4) & \textbf{0.1473} ($\pm$ 0e+0) & 0.1875 ($\pm$ 0e+0) \\
\rowcolor[HTML]{C0C0C0} \cellcolor[HTML]{FFFFFF} 
& \me & \textbf{0.7959} ($\pm$ 5e-4) & \textbf{0.3702} ($\pm$ 3e-4) & \textbf{0.0573} ($\pm$ 7e-3) & 0.8104 ($\pm$ 1e-4) & 0.4415 ($\pm$ 1e-4) & \textbf{0.0055} ($\pm$ 4e-4) & 0.8218 ($\pm$ 3e-4) & 0.1475 ($\pm$ 5e-4) & \textbf{0.0409} ($\pm$ 3e-3) \\
\hline
\multirow{7}{*}{\rotatebox{90}{\textbf{DCN}}}
& Backbone & \textbf{0.7956} ($\pm$ 2e-4) & \textbf{0.3703} ($\pm$ 1e-4) & 1.0000 ($\pm$ 0e+0) & \textbf{0.8106} ($\pm$ 2e-4) & \textbf{0.4413} ($\pm$ 2e-4) & 1.0000 ($\pm$ 0e+0) & \textbf{0.8224} ($\pm$ 2e-4) & 0.1472 ($\pm$ 1e-4) & 1.0000 ($\pm$ 0e+0) \\
& QR-Trick & 0.7901 ($\pm$ 1e-4) & 0.3740 ($\pm$ 1e-4) & 0.5000 ($\pm$ 0e+0) & 0.8078 ($\pm$ 1e-4) & 0.4437 ($\pm$ 1e-4) & 0.5000 ($\pm$ 0e+0) & 0.8094 ($\pm$ 1e-4) & 0.1502 ($\pm$ 2e-5) & 0.5000 ($\pm$ 0e+0) \\
& PEP & 0.7954 ($\pm$ 1e-4) & 0.3705 ($\pm$ 1e-4) & 0.9946 ($\pm$ 1e-5) & 0.8104 ($\pm$ 1e-4) & 0.4414 ($\pm$ 1e-4) & 0.0332 ($\pm$ 5e-4) & 0.8221 ($\pm$ 1e-4) & 0.1475 ($\pm$ 1e-4) & 0.9948 ($\pm$ 1e-4) \\
& OptFS & 0.7954 ($\pm$ 1e-4) & 0.3705 ($\pm$ 1e-4) & 0.9870 ($\pm$ 1e-5) & 0.8105 ($\pm$ 1e-4) & \textbf{0.4413} ($\pm$ 1e-4) & 0.0182 ($\pm$ 1e-4) & 0.8218 ($\pm$ 1e-4) & 0.1475 ($\pm$ 1e-4) & 0.9842 ($\pm$ 2e-4) \\
& ALPT & 0.7954 ($\pm$ 1e-4) & 0.3705 ($\pm$ 1e-4) & 0.2500 ($\pm$ 0e+0) & 0.8104 ($\pm$ 1e-4) & 0.4414 ($\pm$ 1e-4) & 0.2500 ($\pm$ 0e+0) & 0.8215 ($\pm$ 1e-4) & 0.1476 ($\pm$ 3e-5) & 0.2500 ($\pm$ 0e+0) \\
& LSQ+ & 0.7954 ($\pm$ 2e-4) & 0.3704 ($\pm$ 1e-4) & 0.1875 ($\pm$ 0e+0) & 0.8105 ($\pm$ 1e-4) & 0.4415 ($\pm$ 1e-4) & 0.1875 ($\pm$ 0e+0) & \textbf{0.8224} ($\pm$ 1e-4) & 0.1472 ($\pm$ 1e-4) & 0.1875 ($\pm$ 0e+0) \\
\rowcolor[HTML]{C0C0C0} \cellcolor[HTML]{FFFFFF} 
& \me & \textbf{0.7956} ($\pm$ 3e-4) & \textbf{0.3703} ($\pm$ 2e-4) & \textbf{0.0551} ($\pm$ 7e-3) & 0.8105 ($\pm$ 1e-4) & 0.4414 ($\pm$ 1e-4) & \textbf{0.0051} ($\pm$ 5e-4) & \textbf{0.8224} ($\pm$ 3e-4) & \textbf{0.1471} ($\pm$ 3e-4) & \textbf{0.0496} ($\pm$ 3e-3) \\ 
\hline
\multirow{7}{*}{\rotatebox{90}{\textbf{DeepFM}}}
& Backbone & \textbf{0.7956} ($\pm$ 1e-4) & 0.3704 ($\pm$ 1e-4) & 1.0000 ($\pm$ 0e+0) & \textbf{0.8100} ($\pm$ 1e-4) & \textbf{0.4419} ($\pm$ 1e-4) & 1.0000 ($\pm$ 0e+0) & 0.8219 ($\pm$ 3e-4) & 0.1474 ($\pm$ 1e-4) & 1.0000 ($\pm$ 0e+0) \\
& QR-Trick & 0.7899 ($\pm$ 1e-4) & 0.3741 ($\pm$ 4e-5) & 0.5000 ($\pm$ 0e+0) & 0.8068 ($\pm$ 2e-4) & 0.4446 ($\pm$ 2e-4) & 0.5000 ($\pm$ 0e+0) & 0.8080 ($\pm$ 2e-4) & 0.1504 ($\pm$ 1e-4) & 0.5000 ($\pm$ 0e+0) \\
& PEP & 0.7953 ($\pm$ 1e-4) & 0.3705 ($\pm$ 3e-5) & 0.9945 ($\pm$ 1e-4) & 0.8098 ($\pm$ 4e-5) & 0.4421 ($\pm$ 1e-4) & 0.0267 ($\pm$ 7e-4) & 0.8218 ($\pm$ 1e-4) & 0.1474 ($\pm$ 1e-4) & 0.9942 ($\pm$ 1e-4) \\
& OptFS & 0.7954 ($\pm$ 1e-4) & 0.3705 ($\pm$ 1e-4) & 0.9865 ($\pm$ 2e-4) & 0.8075 ($\pm$ 2e-4) & 0.4440 ($\pm$ 2e-4) & 0.0184 ($\pm$ 1e-4) & 0.8213 ($\pm$ 2e-4) & 0.1476 ($\pm$ 1e-4) & 0.9838 ($\pm$ 2e-4) \\
& ALPT & 0.7951 ($\pm$ 1e-4) & 0.3708 ($\pm$ 1e-4) & 0.2500 ($\pm$ 0e+0) & 0.8097 ($\pm$ 3e-5) & 0.4420 ($\pm$ 1e-4) & 0.2500 ($\pm$ 0e+0) & 0.8209 ($\pm$ 1e-4) & 0.1477 ($\pm$ 1e-4) & 0.2500 ($\pm$ 0e+0) \\
& LSQ+ & 0.7955 ($\pm$ 1e-4) & 0.3704 ($\pm$ 3e-5) & 0.1875 ($\pm$ 0e+0) & 0.8099 ($\pm$ 1e-4) & \textbf{0.4419} ($\pm$ 1e-4) & 0.1875 ($\pm$ 0e+0) & 0.8218 ($\pm$ 1e-4) & 0.1474 ($\pm$ 1e-4) & 0.1875 ($\pm$ 0e+0) \\
\rowcolor[HTML]{C0C0C0} \cellcolor[HTML]{FFFFFF} 
& \me & \textbf{0.7956} ($\pm$ 4e-4) & \textbf{0.3702} ($\pm$ 2e-4) & \textbf{0.0573} ($\pm$ 6e-3) & 0.8099 ($\pm$ 1e-4) & \textbf{0.4419} ($\pm$ 1e-4) & \textbf{0.0052} ($\pm$ 3e-4) & \textbf{0.8223} ($\pm$ 3e-4) & \textbf{0.1472} ($\pm$ 1e-4) & \textbf{0.0496} ($\pm$ 2e-3) \\
\hline
\multirow{7}{*}{\rotatebox{90}{\textbf{IPNN}}}
& Backbone & 0.7959 ($\pm$ 1e-4) & 0.3702 ($\pm$ 1e-4) & 1.0000 ($\pm$ 0e+0) & \textbf{0.8110} ($\pm$ 2e-4) & \textbf{0.4409} ($\pm$ 2e-4) & 1.0000 ($\pm$ 0e+0) & \textbf{0.8221} ($\pm$ 2e-4) & \textbf{0.1473} ($\pm$ 4e-5) & 1.0000 ($\pm$ 0e+0) \\
& QR-Trick & 0.7901 ($\pm$ 2e-4) & 0.3740 ($\pm$ 2e-4) & 0.5000 ($\pm$ 0e+0) & 0.8079 ($\pm$ 4e-5) & 0.4436 ($\pm$ 3e-5) & 0.5000 ($\pm$ 0e+0) & 0.8083 ($\pm$ 2e-4) & 0.1503 ($\pm$ 1e-4) & 0.5000 ($\pm$ 0e+0) \\
& PEP & 0.7956 ($\pm$ 1e-4) & 0.3704 ($\pm$ 1e-4) & 0.9945 ($\pm$ 3e-5) & 0.8104 ($\pm$ 2e-5) & 0.4414 ($\pm$ 2e-5) & 0.0451 ($\pm$ 2e-4) & 0.8218 ($\pm$ 3e-4) & 0.1474 ($\pm$ 1e-4) & 0.9946 ($\pm$ 2e-4) \\
& OptFS & 0.7956 ($\pm$ 1e-4) & 0.3704 ($\pm$ 1e-4) & 0.9869 ($\pm$ 1e-4) & 0.8108 ($\pm$ 4e-5) & 0.4411 ($\pm$ 4e-5) & 0.0188 ($\pm$ 1e-4) & 0.8216 ($\pm$ 5e-5) & 0.1475 ($\pm$ 3e-5) & 0.9844 ($\pm$ 1e-4) \\
& ALPT & 0.7957 ($\pm$ 2e-4) & 0.3705 ($\pm$ 1e-4) & 0.2500 ($\pm$ 0e+0) & 0.8108 ($\pm$ 1e-4) & 0.4410 ($\pm$ 1e-4) & 0.2500 ($\pm$ 0e+0) & 0.8213 ($\pm$ 1e-4) & 0.1475 ($\pm$ 5e-5) & 0.2500 ($\pm$ 0e+0) \\
& LSQ+ & 0.7957 ($\pm$ 1e-4) & 0.3703 ($\pm$ 1e-4) & 0.1875 ($\pm$ 0e+0) & 0.8109 ($\pm$ 1e-4) & \textbf{0.4409} ($\pm$ 5e-5) & 0.1875 ($\pm$ 0e+0) & \textbf{0.8221} ($\pm$ 2e-4) & \textbf{0.1473} ($\pm$ 5e-5) & 0.1875 ($\pm$ 0e+0) \\
\rowcolor[HTML]{C0C0C0} \cellcolor[HTML]{FFFFFF} 
& \me & \textbf{0.7960} ($\pm$ 2e-4) & \textbf{0.3700} ($\pm$ 3e-4) & \textbf{0.0609} ($\pm$ 7e-3) & 0.8108 ($\pm$ 1e-4) & 0.4412 ($\pm$ 1e-4) & \textbf{0.0053} ($\pm$ 3e-4) & 0.8220 ($\pm$ 3e-4) & \textbf{0.1473} ($\pm$ 1e-4) & \textbf{0.0412} ($\pm$ 1e-3) \\
\hline
\end{tabular}
}
\end{table*}

%% file: sections/related_work.tex
\section{Related Work}
\subsection{Embedding Compression}
Numerous studies have explored embedding compression techniques in recommendation models. \citet{embed_survey} provide a comprehensive survey of these methods, classifying them into three categories: low-precision, mixed-dimension, and weight-sharing, which target parameter, vector, and matrix-level compression, respectively. Our proposed method search for an ideal precision for each feature embedding, falling under the low-precision category. 

\textbf{Low-precision methods} aim to reduce the precision of embedding parameters through quantization techniques. However, as discussed in Section \ref{sec:introduction}, embedding quantization has received relatively less attention, with most research focusing on low-precision training (LPT) due to its effectiveness in memory compression during training \cite{baidu_lpt, facebook_lpt, alpt}.
\textbf{Mixed-dimension methods} aim to reduce the size of embedding vectors. For example, CpRec \cite{cprec} assigns smaller embedding dimensions to less frequent features, while MDE \cite{mde} adjusts the embedding dimensions according to the size of the associated feature fields. Recent research has utilized neural architecture search (NAS) techniques \cite{nas} to search feature dimensions. For example, ESAPN \cite{esapn} maintains a policy network optimized via reinforcement learning algorithms to decide when to increase embedding dimensions. Further, AutoEmb \cite{autoemb} designs dimension search algorithm through differential architecture search (DARTS) techniques \cite{darts}. 
Additionally, embedding pruning is essentially a mixed-dimension method. PEP \cite{pep} prunes embedding vectors by learning a threshold that dynamically determines the embedding size for different features.
\textbf{Weight-sharing methods} aim to reduce the number of parameters actually used by embedding tables. For example, DoubleHash \cite{double-hash} allows less frequent features to share embeddings. ROBE applies more advanced hash functions to map each embedding parameter into a shared memory space, enhancing compression efficiency. On the other hand, \citet{saec} employs vector quantization to capture the similarity among features. They first cluster the most frequent embeddings to generate a codebook composed of several codewords. Then, each feature embedding is approximately represented by its most similar codeword. Note that vector quantization is essentially the clustering of vectors, which differs from the parameter quantization discussed in this paper.

\subsection{Mixed-Precision Quantization}
In mixed-precision neural networks, prior research has primarily focused on assigning different precision levels across layers to reduce memory consumption and computational overhead \cite{mixed_precision_survey, dnas, haq, hawq, hawq-v3}. They can be categorized into three categories: gradient-based, heuristic-based, and reinforcement learning-based methods. 
\textbf{Gradient-based methods} usually convert the discrete precision selection problem into a continuous one, facilitating optimization through gradient descent algorithms. 
\citet{dnas} propose a differentiable neural architecture search (DNAS) strategy, which utilizes Gumbel Softmax \cite{gumbel} to achieve differentiable precision selection. Subsequent studies, such as BP-NAS \cite{bp-nas}, GMPQ \cite{gmpq}, and SEAM \cite{seam}, have improved upon DNAS. 
\textbf{Heuristic-based methods} develop reasonable sensitivity metrics to assign precision. For example, HAWQ \cite{hawq, hawq-v3} uses Hessian information to guide precision assignment, while Hybrid-Net \cite{hybrid-net} applies principal component analysis (PCA) to identify significant layers and assign higher precision accordingly. 
\textbf{Reinforcement learning-based methods}, like HAQ \cite{haq} and ADRL \cite{adrl}, employ reinforcement learning algorithms to search for optimal precision levels. 
In this paper, we reformulate the precision search as a precision probability distribution learning problem, where the distribution is updated directly via gradient descent, making it a gradient-based method. 

On the other hand, most mixed precision training methods consist of two stages: search and retraining \cite{dnas,hawq,hawq-v3}. The search stage identifies the optimal precision configuration, while the retraining stage usually fine-tunes the model to improve accuracy. However, some studies directly apply mixed-precision quantization to pre-trained models \cite{mpq-ptq, zeroq} or train a mixed-precision network from scratch without retraining \cite{retrain-free}. \me\ requires retraining, and the exploration of retraining-free mixed-precision embeddings is left for future work.

%% file: sections/conclusion.tex
\section{Conclusion}
In this paper, we identify a critical drawback of quantization-aware training for embedding compression, that is its inability to distinguish feature importance. To this end, we propose the mixed-precision embeddings (\me) algorithm to enhance the compression efficiency, which identifies an appropriate precision for each feature to balance model accuracy and memory usage. Specifically, we first group features by frequency to simplify the search space for precision and then learn a probability distribution over precision levels for each group. \me\ will sample the final bit-widths based on the optimized probability distribution and then retrain the model to achieve better accuracy. 
Extensive experiments are conducted to evaluate the performance of \me. The results demonstrate that the \me\ significantly outperforms the state-of-the-art methods, achieving a 200-fold compression without any loss in prediction accuracy on the Criteo dataset. Additionally, \me\ is implemented as a plug-in embedding layer module based on PyTorch, ensuring ease of use.